\documentclass[12pt,fleqn,twoside]{article}
\usepackage{espcrc1}

\usepackage{graphicx}
\usepackage[figuresright]{rotating}

\newcommand{\AmS}{{\protect\the\textfont2
  A\kern-.1667em\lower.5ex\hbox{M}\kern-.125emS}}

\title{
\thispagestyle{empty}
\vspace{-25mm}
\rightline{\small DESY 04-005~~~~~}
\rightline{\small ITEP-LAT/2004-01~~~~~}
\rightline{\small KANAZAWA-04-02~~~~~}
\vspace{10mm}
Profiles of the broken string in two-flavor QCD below and above the finite temperature transition.
}
\author{
V. G. Bornyakov\address{Institute for High Energy Physics, RU-142284 Protvino, Russia\\[-0.5em]},
M. N. Chernodub\address[ITEP]{ITEP, B.Cheremushkinskaya 25, RU-117259 Moscow, Russia\\[-0.5em]},
H. Ichie\address[KITP]{Institute for Theoretical Physics, Kanazawa University, Kanazawa 920-1192, Japan\\[-0.5em]},
Y. Koma\address[MPI]{Max-Planck-Institut f\"ur Physik, D-80805 M\"unchen, Germany\\[-0.5em]},
Y. Mori$^{\rm c}$,
M. I. Polikarpov$^{\rm b}$,
G. Schierholz\address[DESY]{NIC/DESY Zeuthen, Platanenallee 6, D-15738 Zeuthen,\\Deutsches Elektronen-Synchrotron DESY D-22603 Hamburg, Germany\\[-0.5em]},
 H.~St\"uben\address{Konrad-Zuse-Zentrum f\"ur Informationstechnik
Berlin, D-14195 Berlin, Germany\\[-0.5em]} and
T.~Suzuki$^{\rm c}$
}

\begin{document}

\maketitle

\begin{abstract}

We study the Abelian flux tube profile of mesonic and the baryonic configurations of static quarks below and above the finite-temperature transition in $N_f=2$ full QCD.
To reduce effects of ultra-violet fluctuations we measure Abelian distributions of the action density, the color-electric field and the monopole current after fixing to the maximally Abelian gauge.
Changing the distance $R$ between quarks for fixed $T/T_c<1$, one can see a clear signal of string breaking for large $R$.
The electric field becomes Coulomb-like and the circulating monopole current disappears.
We study also the temperature dependence of the flux profile for fixed $R \approx 0.7$ fm.
The disappearance of the squeezed flux is observed clearly above $T_c$.
Similar behaviors are observed both in the mesonic and the baryonic systems.
\end{abstract}

\section{Introduction}
The study of various distributions inside the flux tubes formed in meson and baryon systems is important to understand the internal structure of mesons and baryons and to learn more about the QCD vacuum properties.
In full QCD the flux tube, which is usually modeled by the bosonic string, is expected to disappear (string breaking) when the distance between quarks becomes large enough.
How the flux tube evolves and how it disappears when the distance between quarks or the temperature increase beyond respective critical values are still open questions to be addressed in this paper.

In zero temperature full QCD no clear evidence for string breaking has been found so far \cite{Aoki:1998sb,Bolder:2000un,DIK_2003_ZERO}.
This is due to poor overlap of the Wilson loop with the two-meson channels \cite{Bernard:2001tz}.
At zero temperature the flux tube has been extensively studied in quenched QCD \cite{Bali_1994} while there were only few lattice studies
\cite{DIK_2003_ZERO,Ichie:2002mi,Ichie:2003} in full QCD.
In the latter papers the calculations were done in the Abelian projected theory after fixing to the maximally Abelian gauge (MAG).
The use of Abelian projected quantities allowed us to reduce ultra-violet fluctuations and improve the signal to noise ratio in comparison with the non-Abelian quantities.
In \cite{DIK_2003_ZERO} the profile of the Abelian flux tube between a quark and an anti-quark of the length $R\sim 1.6$~fm was presented.
The three quark system has been studied in \cite{Ichie:2002mi,Ichie:2003}.
It has been shown that the baryonic flux tube has a shape consistent with the so-called Y-type flux tube in agreement with other numerical results \cite{Takahashi_2002}, and theoretical predictions \cite{Kamizawa:1992hb,Chernodub:1998ie,Kuzmenko:2000bq,Koma_2001_02}.
In finite temperature full QCD string breaking is much easier to observe using Polyakov loop correlators \cite{DeTar:1998qa,DIK_2003}.
The state created by Polyakov loops has considerable overlap with both the string and the broken string states.

In this paper, we study flux tube profiles in the mesonic and the baryonic systems in finite temperature lattice QCD with dynamical fermions at distances below and above the string breaking distance.
In \cite{DIK_2003} we have studied string breaking in this theory using Abelian Polyakov loop correlators.
In that paper the transition temperature was also determined.
Following our previous work \cite{DIK_2003_ZERO,Ichie:2003,DIK_2003} we investigate the structure of the flux tube with the help of Abelian observables after fixing to MAG.
In particular, we study the monopole and the photon parts of Abelian flux tube profiles.
This permits us to get good signal for both mesonic and baryonic flux tubes of large size and to find further confirmation of the dual superconductor mechanism of confinement.

\begin{table}[btp]
\begin{center}
\begin{tabular}{|c|c|c|r||c|c|c|r|}
\hline
\multicolumn{4}{|c|}{$\beta=5.2$}&\multicolumn{4}{|c|}{$\beta=5.25$} \\
\hline
 $\kappa$ & $T/T_{c}$ &Trajectories & $\tau_{int}$ & $\kappa$ & $T/T_{c}$ &Trajectories & $\tau_{int}$\\
 \hline
 0.1330 & 0.80 & 7129  & 24 & 0.1335  & 0.92 & 7500  & 90  \\
 0.1335 & 0.87 & 4500  & 54 & 0.13375 & 0.96 & 9225  & 200 \\
 0.1340 & 0.94 & 3000  & 62 & 0.1339  & 0.97 & 12470 & 440 \\
 0.1344 & 1.00 & 8825  & 520&         & &       &     \\
 0.1360 & 1.28 & 3699  & 46 &         & &       &     \\
 \hline
\end{tabular}
\end{center}
~\\[-10mm]
\caption{
Parameters and statistics of the simulation, together with the integrated
autocorrelation time $\tau_{int}$. The length of the trajectory is
$\tau = 0.25$.
}
~\\[-10mm]
\label{tab1}
\end{table}

\section{Simulation details}

To study QCD with dynamical quarks we consider $N_{f}=2$ flavors of degenerate quarks, using the Wilson gauge field action and non-perturbatively  ${\cal O}(a)$ improved Wilson fermions~\cite{Booth:2001qp}.
Configurations are generated on  $16^3\times 8$ lattice at $\beta=5.2$, $0.1330\le \kappa \le 0.1360$, corresponding to temperatures below and above the transition temperature $T_{c}=213(10)$~MeV ($\kappa_t=0.1344$) and at $\beta=5.25$ for $\kappa < \kappa_t$~\cite{DIK_2003}.
Details of the simulations can be found in~\cite{DIK_2003}.
Table~\ref{tab1} shows the number of configurations used in this study.
We perform the MAG fixing on generated configurations employing the simulated annealing algorithm \cite{Bali:1996dm}.
The Abelian projection procedure \cite{MaA} defines the diagonal link matrices
\begin{eqnarray}
u_{\mu}(s)&=&\mbox{diag} \{u^{1}_{\mu}(s), u^{2}_{\mu}(s),u^{3}_{\mu}(s)\},\\
u^{a}_{\mu}(s)&=&e^{i\theta^{a}_{\mu}(s)}~,\label{eq:UAbel}
\end{eqnarray}
where $\theta^{a}_{\mu}(s)$ can in turn be decomposed into monopole (singular) and photon (regular) parts~\cite{ploop,Smit_1991,Suzuki_1995}:
\begin{eqnarray}
\theta^{a}_{\mu}(s) = \theta^{{\rm mon},a}_{\mu}(s) +\theta^{{\rm ph},a}_{\mu}(s).
\label{eq:decomposition}
\end{eqnarray}
Monopole and photon parts are defined up to irrelevant terms by following expressions:
\begin{eqnarray}
\theta^{{\rm mon},a}_{\mu}(s)&=&-2 \pi \sum_{s'}  D(s-s')\, \partial_{\nu}'\, n^{a}_{\nu\mu}(s)~,\label{eq:theta:mon}\\
\theta^{{\rm ph},a}_{\mu}(s)&=&-\sum_{s'}  D(s-s')\, \partial_{\nu}'\, \overline{\theta}^{a}_{\nu\mu}(s)~,\label{eq:theta:ph}
\end{eqnarray}
where
\begin{equation}
\bar{\theta}^{a}_{\mu\nu}(s) = \partial_{\mu} \theta^{a}_{\nu}(s) -
 \partial_{\nu} \theta^{a}_{\mu}(s) - 2\pi n^{a}_{\mu\nu}(s),\,\,
\bar{\theta}^{a}_{\mu\nu}(s) \in (-\pi,\pi],\,\,
n^{a}_{\mu\nu}(s) = 0, \pm 1, \pm 2 \,,
\end{equation}
$\partial_{\nu}\,'$ and $\partial_{\nu}$ are backward and forward lattice derivatives, respectively, and $D(s)$
denotes the lattice Coulomb propagator.

We study the Abelian action density
\begin{equation}
\rho_{\rm ab}(s) = \frac{\beta}{3} \sum_{\mu>\nu}\sum_{a}
{\rm cos}(\bar{\theta}^{a}_{\mu\nu}(s))\, ,
\label{action}
\end{equation}
the Abelian color-electric field
\begin{equation}
E^{a}_{j}(s)=i\bar{\theta}^{a}_{j4}(s),
\end{equation}
and the monopole current
\begin{equation}
k^{a}_{\mu}(s)=-\frac{i}{4\pi}\epsilon_{\mu\nu\rho\sigma}\partial_{\nu}\bar{\theta}^{a}_{\rho\sigma}(s+\hat{\mu})\,.
\end{equation}
We consider three types of Polyakov loops to create static sources: \\
Abelian:
\begin{equation}
L^{a}_{\rm ab}(\vec{s}) = \exp \left\{{\rm i} \sum_{t=1}^{N_t} \theta^{a}_{4}(\vec{s},t) \right\}\,,\,\,\,\, L_{\rm ab}(\vec{s})=\frac{1}{3}\sum_{a}^{3}
L^{a}_{\rm ab}(\vec{s})\,;\label{eq:PL_op_Ab}
\end{equation}
monopole:
\begin{equation}
L^{a}_{\rm mon}(\vec{s})= \exp \left\{{\rm i} \sum_{t=1}^{N_t} \theta^{{\rm mon},a}_{4}(\vec{s},t) \right\}\,,\,\,\,\, L_{\rm mon}(\vec{s})=
\frac{1}{3}\sum_{a}^{3} L^{a}_{\rm mon}(\vec{s})\,;\label{eq:PL_op_Mo}
\end{equation}
photon:
\begin{equation}
L^{a}_{\rm ph}(\vec{s})= \exp \left\{{\rm i} \sum_{t=1}^{N_t} \theta^{{\rm Ph},a}_{4}(\vec{s},t) \right\}\,,\,\,\,\, L_{\rm ph}(\vec{s})=\frac{1}{3}\sum_{a}^{3}
L^{a}_{\rm ph}(\vec{s})\,.\label{eq:PL_op_Ph}
\end{equation}
The vacuum averages of our observables are defined for the mesonic case by
\begin{eqnarray}
\langle \rho_{\rm ab}(s) \rangle_{Q\bar{Q}} = \frac{\langle \rho_{\rm ab}(s) {\cal P}_{Q\bar{Q}}(r) \rangle}{\langle {\cal P}_{Q\bar{Q}}(r) \rangle} - \langle \rho_{\rm ab}(s)\rangle,
\label{eq:Oper-def_AD}
\end{eqnarray}
\begin{equation}
\langle E_{j}(s) \rangle_{Q\bar{Q}} = \frac{\langle \frac{1}{3} \sum_{a}
E^{a}_{j}(s) L^a(\vec{s}_1)L^a(\vec{s}_2)^*\rangle}{\langle {\cal P}_{Q\overline{Q}}(r) \rangle}\, ,
\label{QQbar_elfield}
\end{equation}
\begin{equation}
\langle k_{j}(s) \rangle_{Q\bar{Q}} = \frac{\langle \frac{1}{3} \sum_{a}
k^{a}_{j}(s) L^a(\vec{s}_1)L^a(\vec{s}_2)^*\rangle}{\langle {\cal P}_{Q\overline{Q}}(r) \rangle}\,
\label{QQbar_mcurrent}
\end{equation}
and for the baryonic case by
\begin{eqnarray}
\langle \rho_{\rm ab}(s) \rangle_{3Q} = \frac{\langle \rho_{\rm ab}(s)
{\cal P}_{3Q}(r_{\rm Y}) \rangle}{\langle {\cal P}_{3Q}(r_{\rm Y}) \rangle} - \langle \rho_{\rm ab}(s)\rangle,
\label{eq:Oper-def_AD2}
\end{eqnarray}
\begin{equation}
\langle E_{j}(s) \rangle_{3Q} = \frac{\langle \frac{1}{3!}\sum
|\varepsilon_{abc}| E^{a}_{j}(s) L^a(\vec{s}_1)L^b(\vec{s}_2)L^c(\vec{s}_3)
\rangle} {\langle {\cal P}_{3Q}(r_{\rm Y}) \rangle}\, ,
\label{QQQ_elfield}
\end{equation}
\begin{equation}
\langle k_{j}(s) \rangle_{3Q} = \frac{\langle \frac{1}{3!}\sum
|\varepsilon_{abc}| k^{a}_{j}(s) L^a(\vec{s}_1)L^b(\vec{s}_2)L^c(\vec{s}_3)
\rangle} {\langle {\cal P}_{3Q}(r_{\rm Y}) \rangle}~,
\label{QQQ_mcurrent}
\end{equation}
\begin{figure}[!tpb]
\begin{center}
\includegraphics[width=120mm]{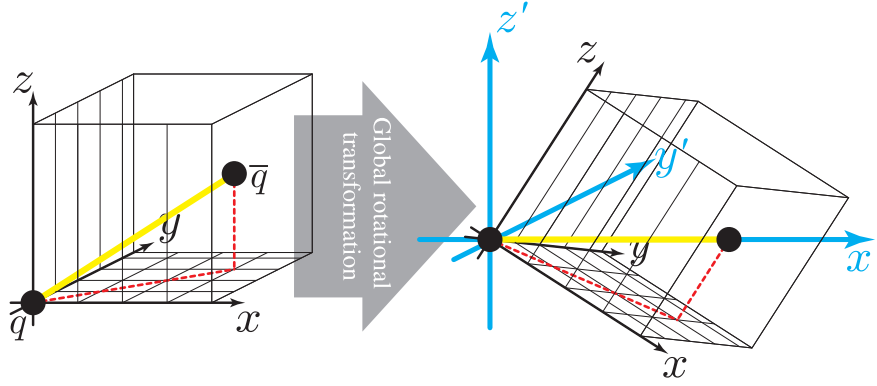}
\end{center}
\vspace{-12mm}
\caption{
Illustration of a global rotational transformation used for off-axis distances in the $Q\bar{Q}$ system.
}
\label{fig:off-axis}
\vspace{-6mm}
\end{figure}
where
\begin{equation}
{\cal P}_{Q\overline{Q}}(r)=\frac{1}{3} \sum_{a=1}^3\,L^{a}(\vec{s}_1)L^{a}(\vec{s}_2)^{*}\, ,\label{eq:Source_QQbar}
\end{equation}
\begin{equation}
{\cal P}_{3Q}(r_{\rm Y})=\frac{1}{3!} \sum_{a,b,c}|\varepsilon_{abc}|\,L^{a}(\vec{s}_1)L^{b}(\vec{s}_2)L^{c}(\vec{s}_3)\, ,\label{eq:Source_3Q}
\end{equation}
$\vec{s}_i$ denotes the position of the $i^{\rm th}$ quark,
$r$ is the distance between quark and anti-quark
\begin{equation}
r=|\vec{s}_1-\vec{s}_2|\, ,
\label{eq:R_QQbar}
\end{equation}
and $r_{\rm Y}$ is the minimal Y-type distance between the three quarks, {\it i.e.} the sum of the distances from the three quarks to the Fermat point,
\begin{equation}
r_{\rm Y}^2=\frac{1}{2}\sum_{i>j}\,r_{ij}^2+2\sqrt{3}S_{\Delta}\,,
\label{eq:R_min}
\end{equation}
$r_{ij}=|{\vec s}_i- {\vec s}_j|$, $S_{\Delta}$  is the area of the corresponding triangle.
Eq.(\ref{eq:R_min}) defines $r_{\rm Y}$ when all angles in the three quarks triangle are less than $2\pi/3$.
If one of the angles is equal or larger than $2\pi/3$, then~\cite{Takahashi_2002} $r_{\rm Y}=\sum_{i>j}r_{ij} - \max r_{ij}$.
$L^{a}(\vec{s})$ stands for one of three types of Polyakov loops introduced above.
The lattice distance is converted into the physical scale using the lattice spacing $a(\beta)$, i.e., $R=ra$ and $R_{\rm Y}=r_{\rm Y}a$.

For mesonic system we need both on-axis and off-axis distances.
In the latter case the x-axis was rotated along the line connecting the sources as shown in Figure \ref{fig:off-axis}.

\section{Mesonic system}
\subsection{The finite temperature transition and the static potential}
~\\[-8mm]

\begin{figure}[tpb]
\begin{center}
\includegraphics[width=79mm,clip=false]{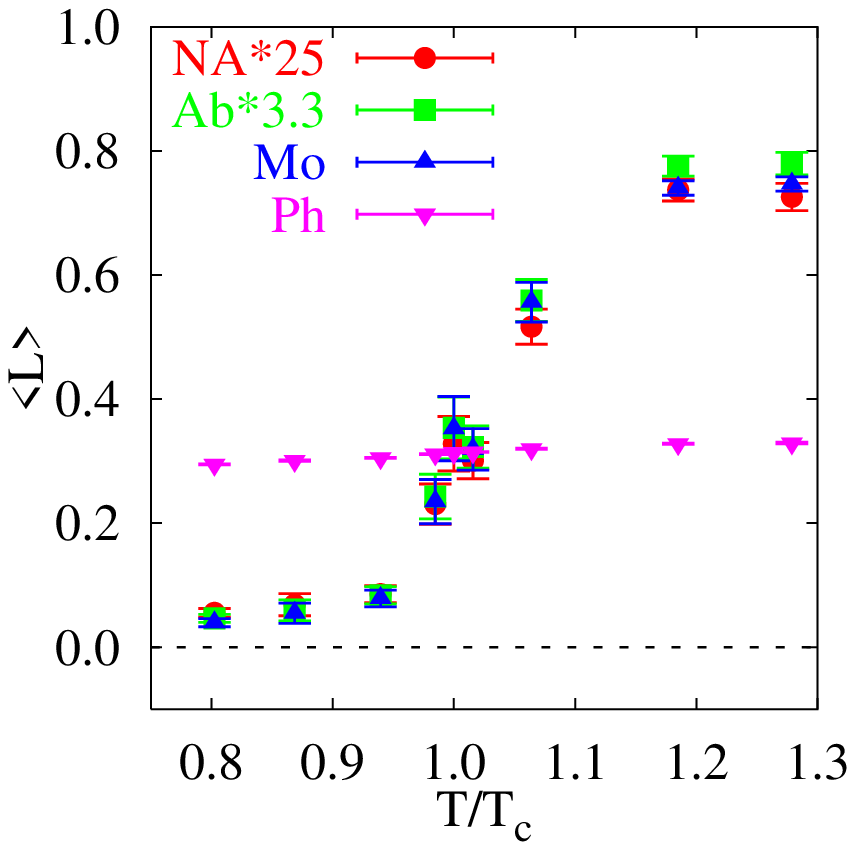}
\includegraphics[width=76mm,clip=false]{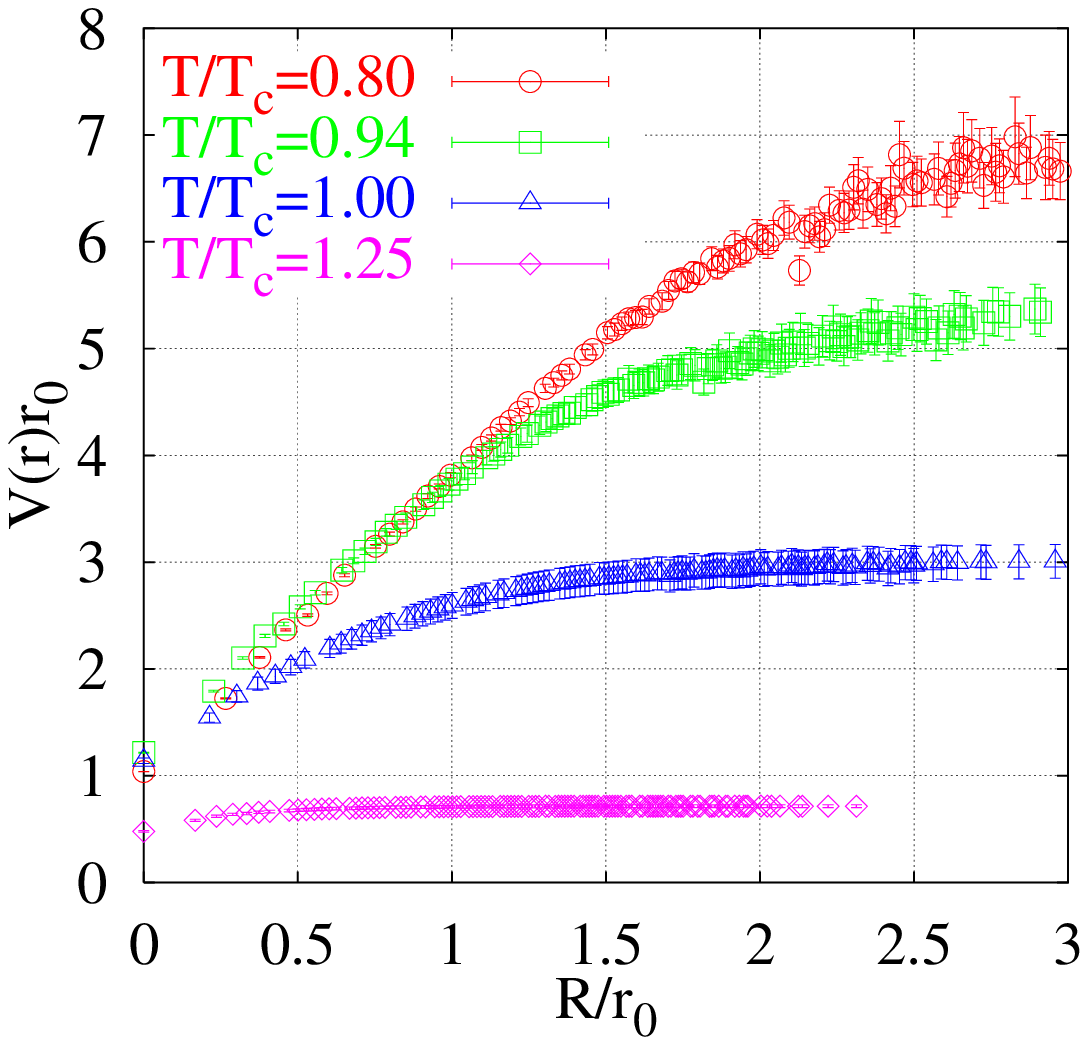}
\end{center}
\vspace{-15mm}
\caption{
Expectation values of the non-Abelian~("NA"$\times 25$), the Abelian~("Ab"$\times$ 3.3), the monopole ("Mo") and the photon ("Ph") Polyakov loops (left) and the static potential from monopole Polyakov loop correlators (right) for $\beta=5.2$.
}
\label{fig:de-confinement_phase_transition}
\end{figure}

Some features of the finite-temperature full QCD have been studied in our previous paper \cite{DIK_2003}.
We show expectation values of various Polyakov loops in  Figure \ref{fig:de-confinement_phase_transition}~(left).
One can see that all Polyakov loops are smooth functions of the temperature and that Abelian and monopole Polyakov loop behaviors are qualitatively the same as that of the non-Abelian Polyakov loop, while the photon Polyakov loop is almost constant across the transition~\cite{DIK_2003}.
The transition is a smooth one (cross-over) in contrast with the first-order phase transition in
quenched QCD.

The heavy quark potential $V(r,T)$ is determined, up to an entropy contribution, from the Polyakov loop correlator:
\begin{equation}
\frac{1}{T} V(r,T)= - \ln \left\langle L(\vec{s}) L^{\dagger}(\vec{s}\,')\right\rangle\,.
\label{hqp}
\end{equation}
In \cite{DIK_2003} we measured non-Abelian, Abelian, monopole and photon potentials using for $ L(\vec{s})$ respective Polyakov loop operators.
When  $r\to \infty$,
\begin{equation}
\langle L(\vec{s}) L^{\dagger}(\vec{s}\,')\rangle \longrightarrow |\langle L\rangle|^2\,,
\end{equation}
where $|\langle L\rangle|^2 \neq 0$, since the global $Z_3$ is broken by fermions.

In Figure \ref{fig:de-confinement_phase_transition} (right)  the monopole part
of the heavy quark potential is shown.
We have analyzed the static potential assuming that at temperatures ~$T < T_c$ the Polyakov loop correlator can be described in terms of the string and two heavy-light meson states.
Then
\begin{equation}
\langle L(\vec{s}) L^{\dagger}(\vec{s}\,') \rangle = e^{-(V_0(T)+V_{\mathrm{string}}(r,T))/T} + e^{-2E(T) /T}\,,
\label{eq:two:exp}
\end{equation}
where for the non-Abelian potential $V_{\rm string}(r,T)$ is given in Ref.~\cite{Gao:kg}, while for the monopole potential we used the linear term only.
The energy $E(T)$ can be written as
\begin{equation}
 E(T)=\frac{1}{2}V_0 + m(T)\,,
\label{meff}
\end{equation}
where $m(T)$ can be considered as a constituent quark mass \cite{Satz:2001kf}, $V_0$ is a self-energy of the static source.
The second term of Eq(\ref{meff}) gives rise to the flattening of the
potential at large distances and describes the effect of creation of a pair
of constituent quarks.
\begin{figure}[tpb]
\begin{center}
\includegraphics[width=150mm]{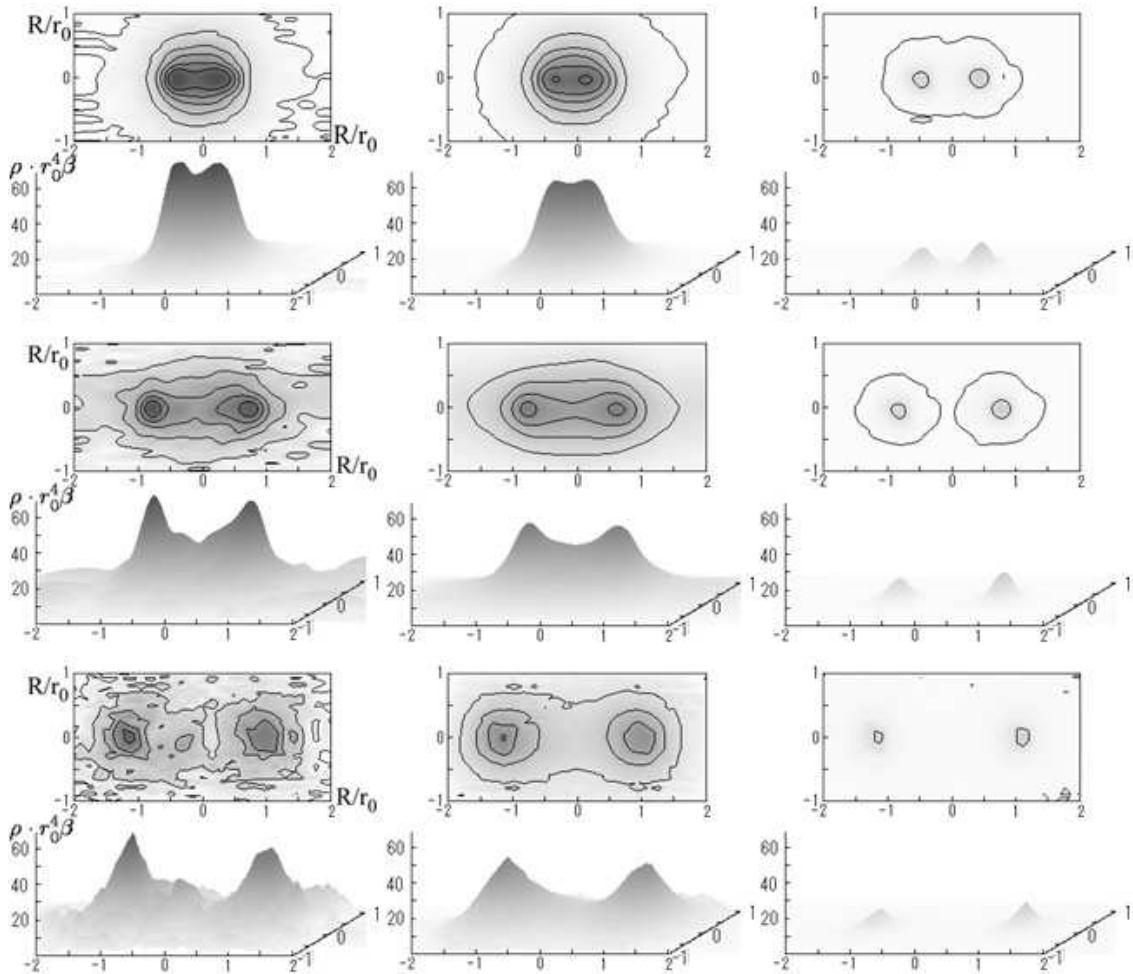}
\end{center}
~\\[-20mm]
\caption{
The profile of the action density of the mesonic system at $T/T_{c}=0.94$.
The sources are made of Abelian (left), monopole (center) and photon (right) Polyakov loops.
The $Q\bar{Q}$ distances are $R/r_{0}=0.91$ (top), 1.59 (middle), 2.25 (bottom).
}
\vspace{-5mm}
\label{fig:Action_density_MESON}
\end{figure}

\subsection{Profiles of the mesonic system}
~\\[-8mm]

As we discussed in the Introduction, in full QCD the flux tube disappears
at large distances due to the light quark-anti-quark pair creation.
In this subsection we demonstrate how this happens for the mesonic flux tube when the $Q-\bar{Q}$ distance $R$ is increased at fixed temperature $T$.
We measured profiles of the action density, the color-electric field and the monopole current at $T/T_c=0.94$  for $R/r_{0}=0.91$, 1.59 and 2.25.
The static quark-anti-quark pair was created by Abelian, monopole, or photon Polyakov loops.
\begin{figure}[tpb]
\begin{center}
\includegraphics[width=50mm]{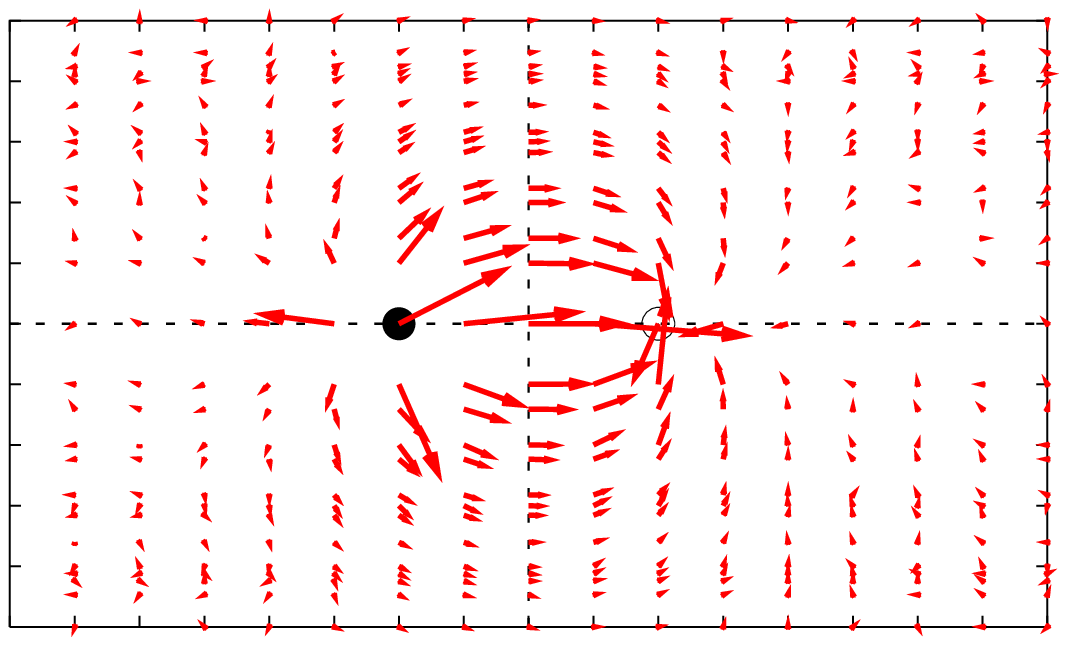}
\includegraphics[width=50mm]{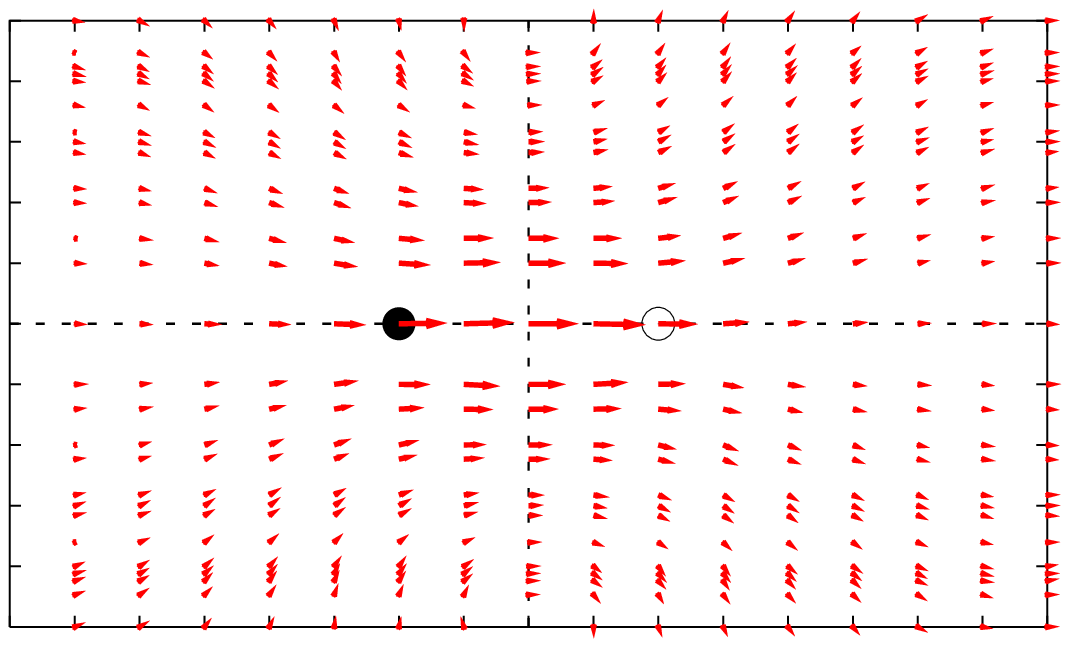}
\includegraphics[width=50mm]{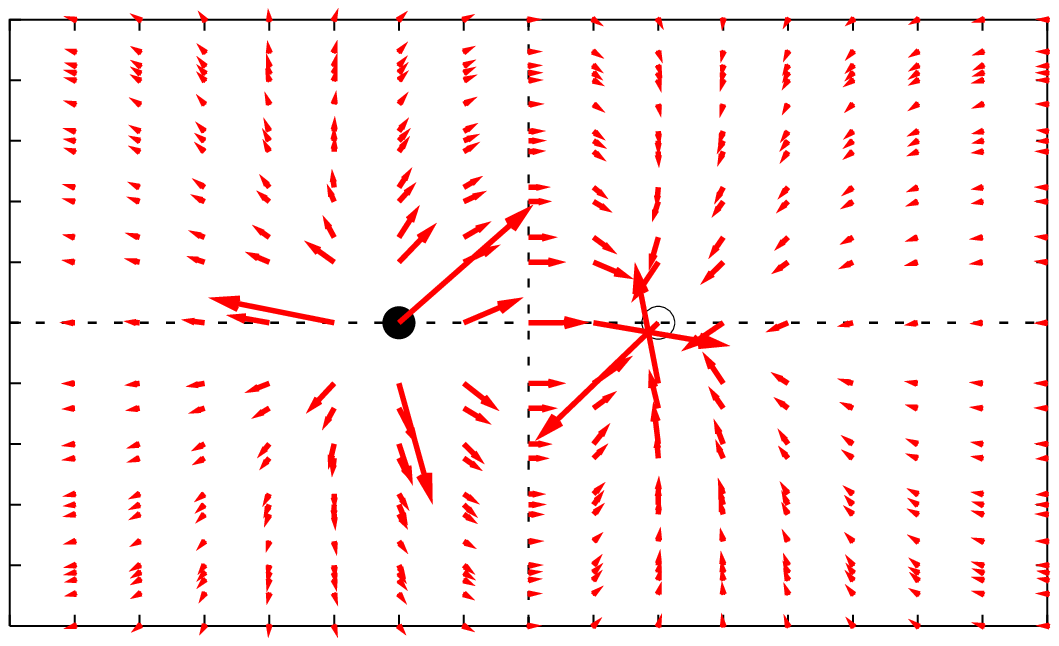}\\
\includegraphics[width=50mm]{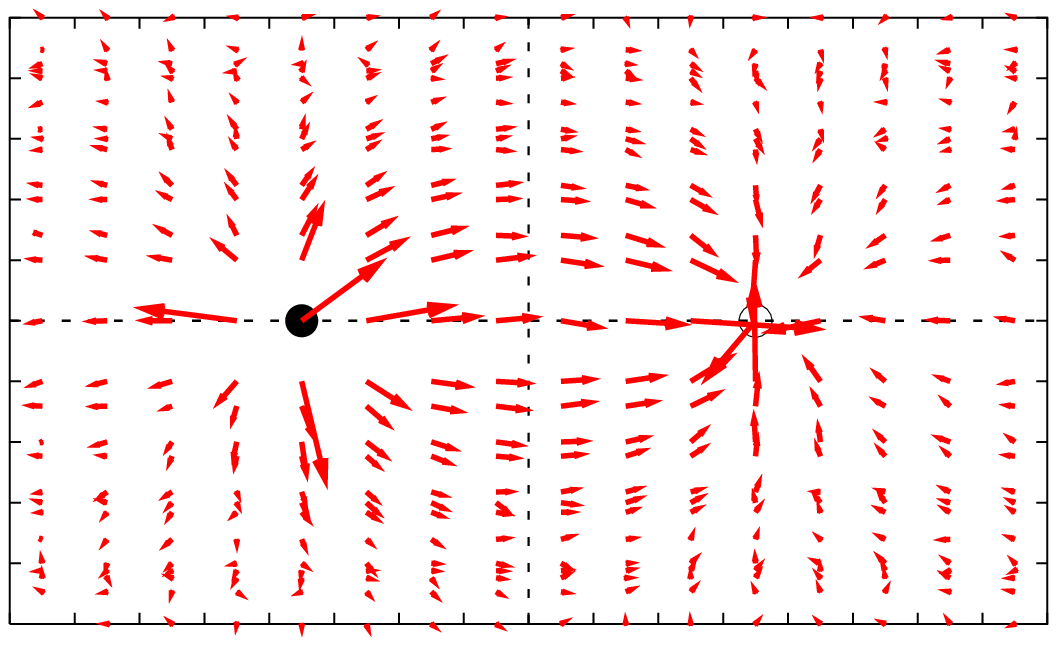}
\includegraphics[width=50mm]{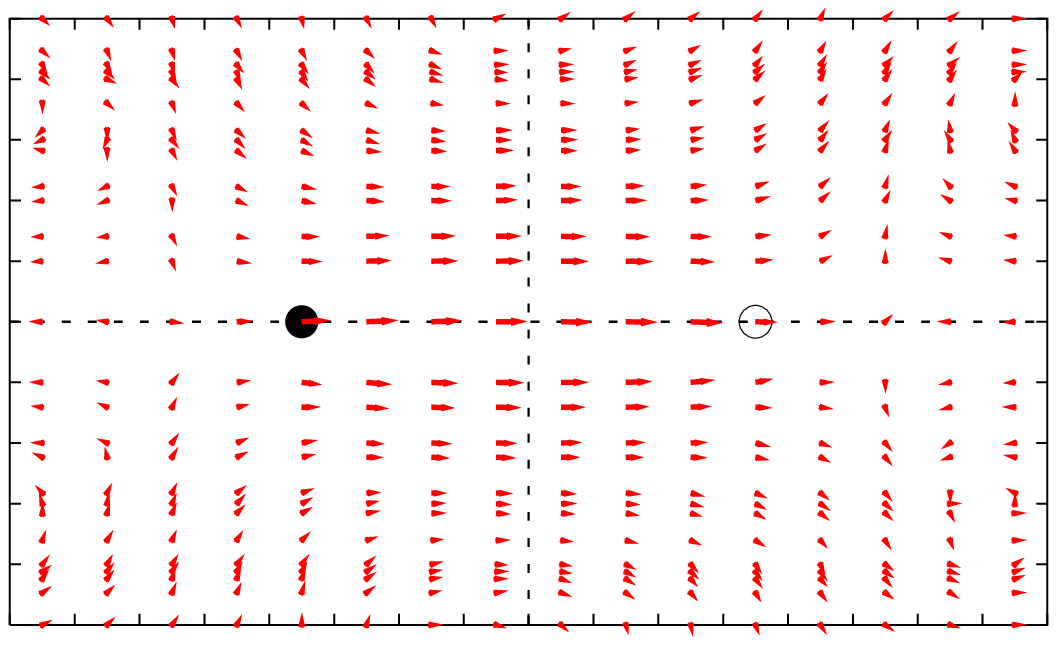}
\includegraphics[width=50mm]{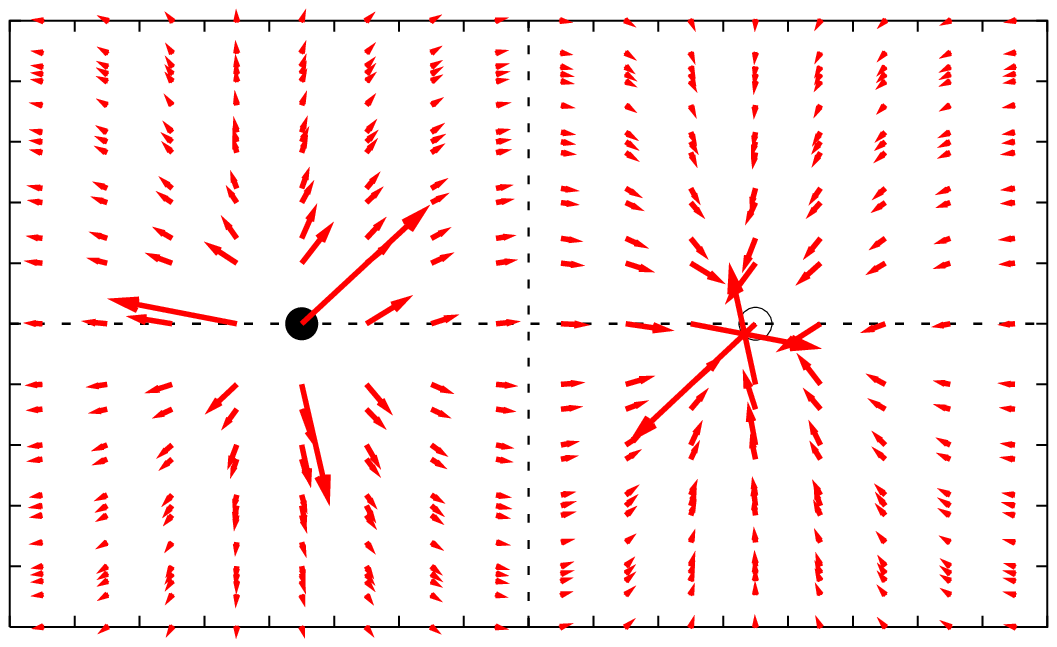}\\
\includegraphics[width=50mm]{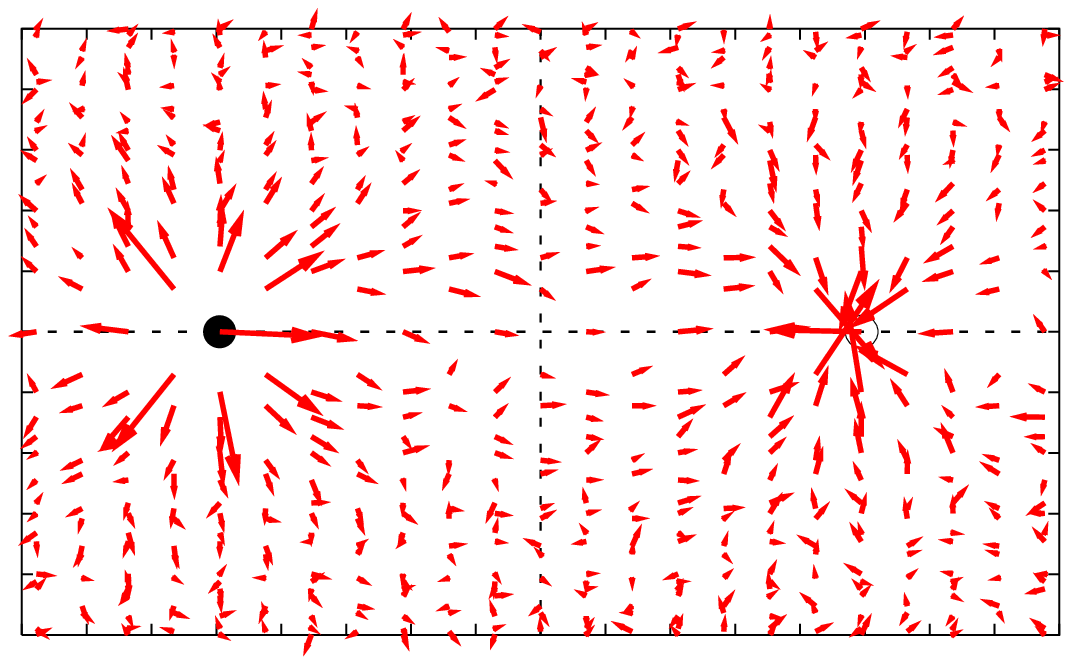}
\includegraphics[width=50mm]{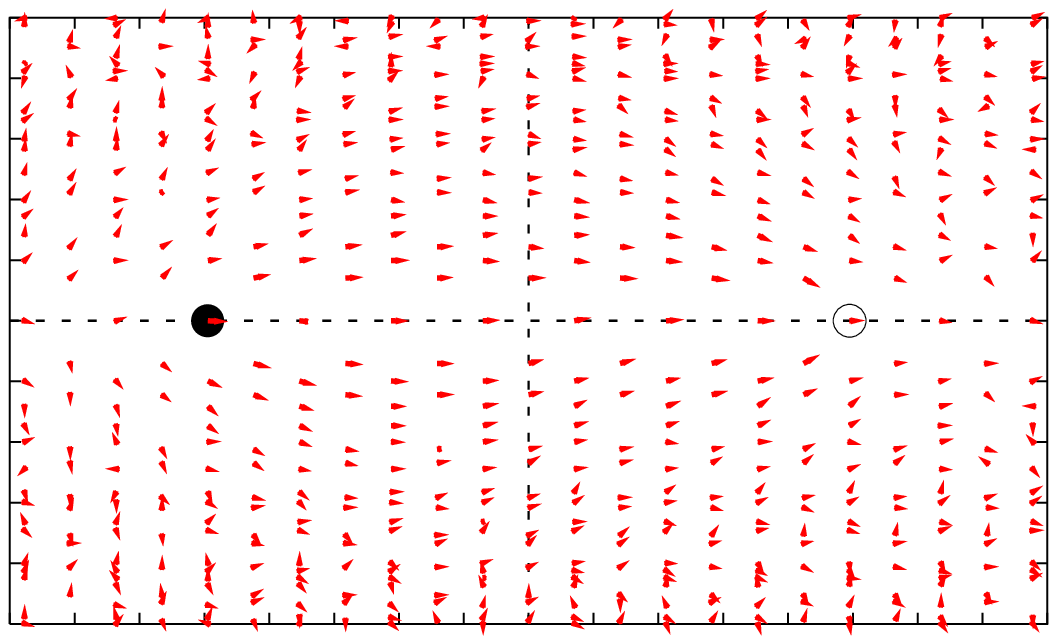}
\includegraphics[width=50mm]{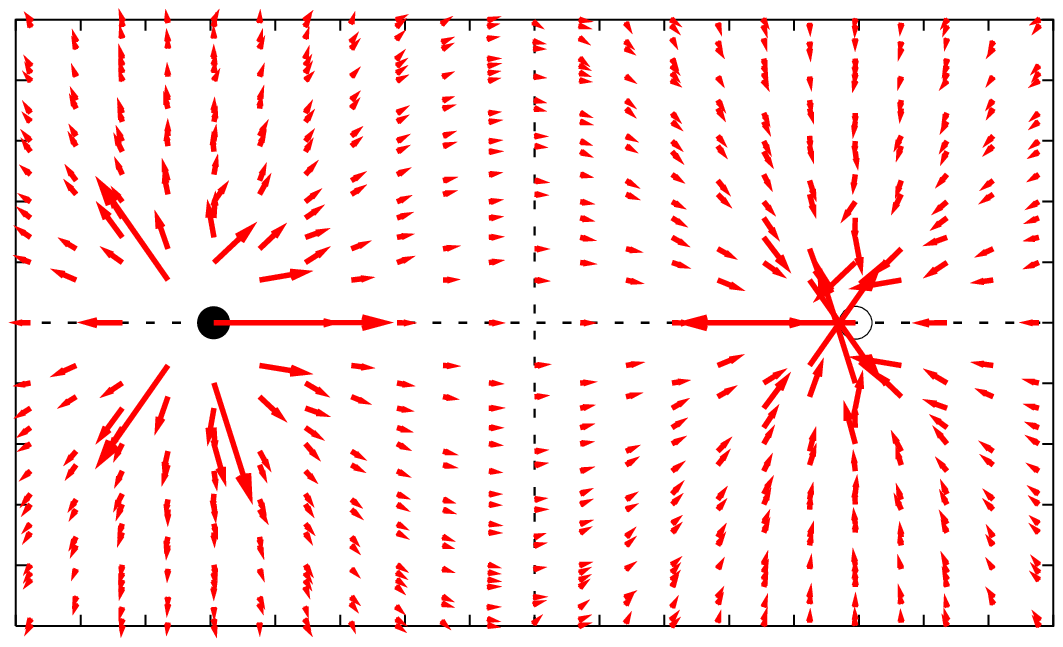}\\
\end{center}
\vspace{-15mm}
\caption{
The profile of the color-electric field for the mesonic flux tube at $T/T_{c}=0.94$.
The sources and distances are as in Figure \ref{fig:Action_density_MESON}.
}
\vspace{-5mm}
\label{fig:Electric_field_MESON}
\end{figure}

Figure \ref{fig:Action_density_MESON} shows profiles of the action density.
We show both the two-dimensional contour plot and the three-dimensional plot for every distance.
In the contour plots, the horizontal axis is always along the line connecting the static sources.
First two columns of this figure indicate that the flux tube persists for distances $R/r_{0}=0.91$, 1.59 while for $R/r_{0}=2.25$ it seems to disappear, leaving only lumps of the action density around the sources.
Note that the latter distance is roughly the distance where the static potential, shown in Figure  \ref{fig:de-confinement_phase_transition}, becomes flat.
The difference between two columns is that in the monopole part the data has much lower noise and two-peaks at locations of the quark and anti-quark are much less pronounced.
In the photon part only two lumps around sources can be seen at all distances.

\begin{figure}[tpb]
\begin{center}
\includegraphics[width=45mm]{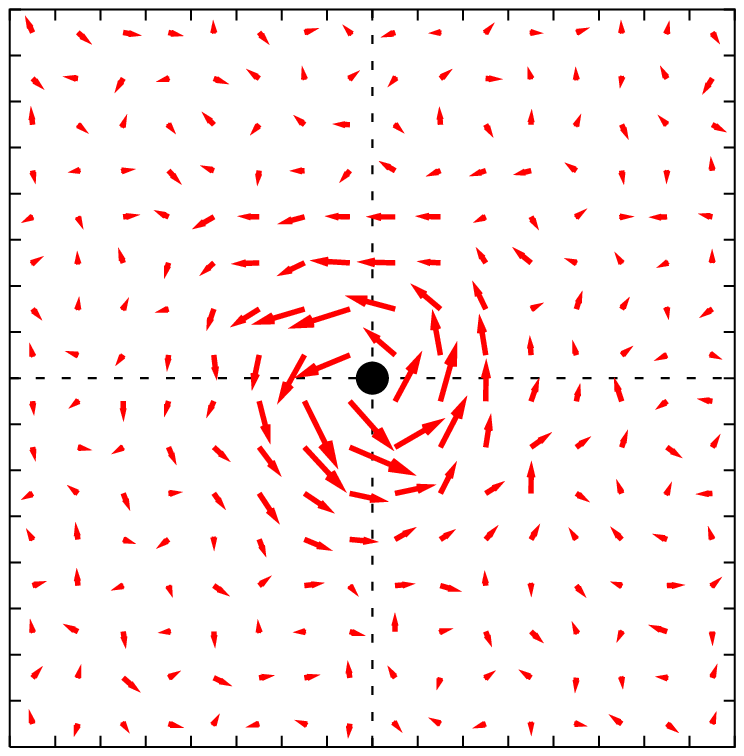}
\includegraphics[width=45mm]{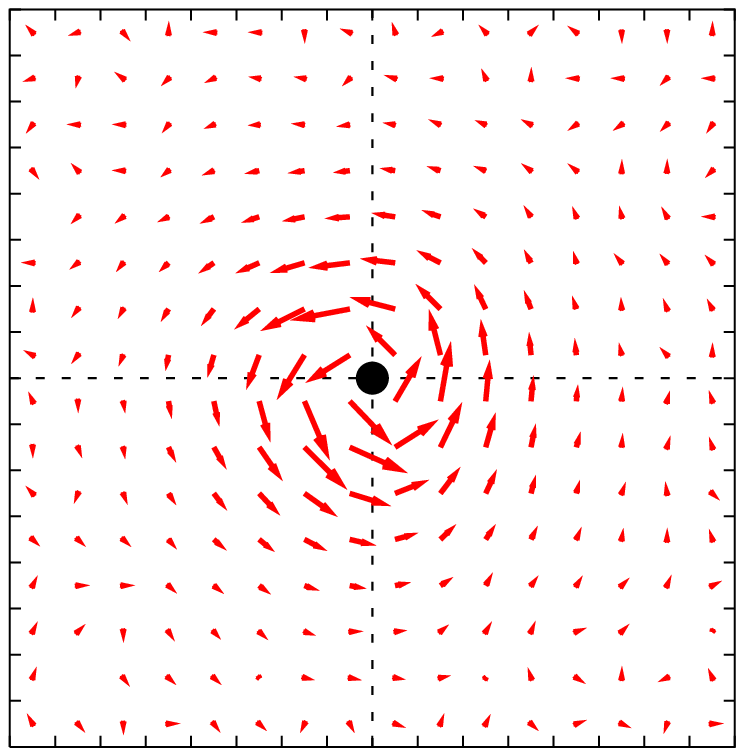}
\includegraphics[width=45mm]{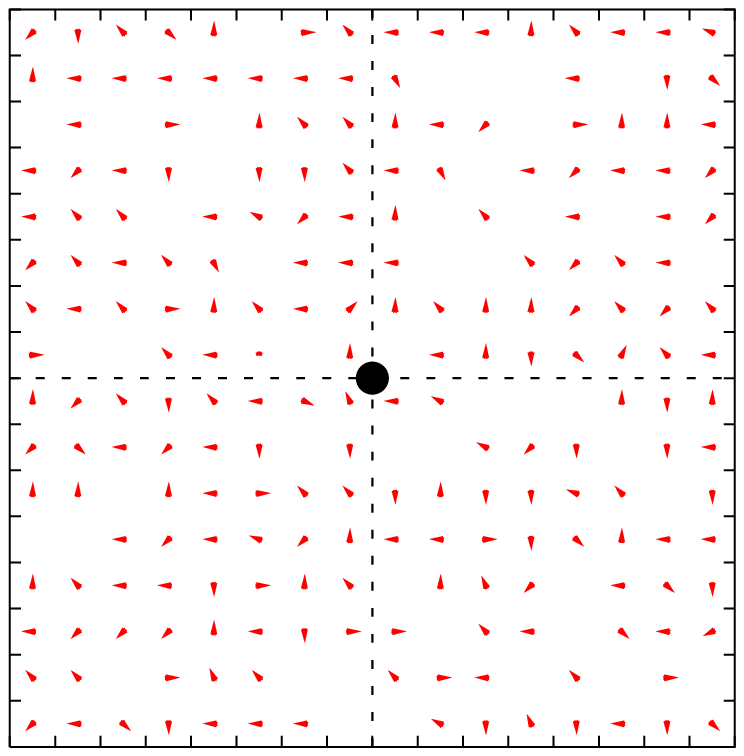}\\
\includegraphics[width=45mm]{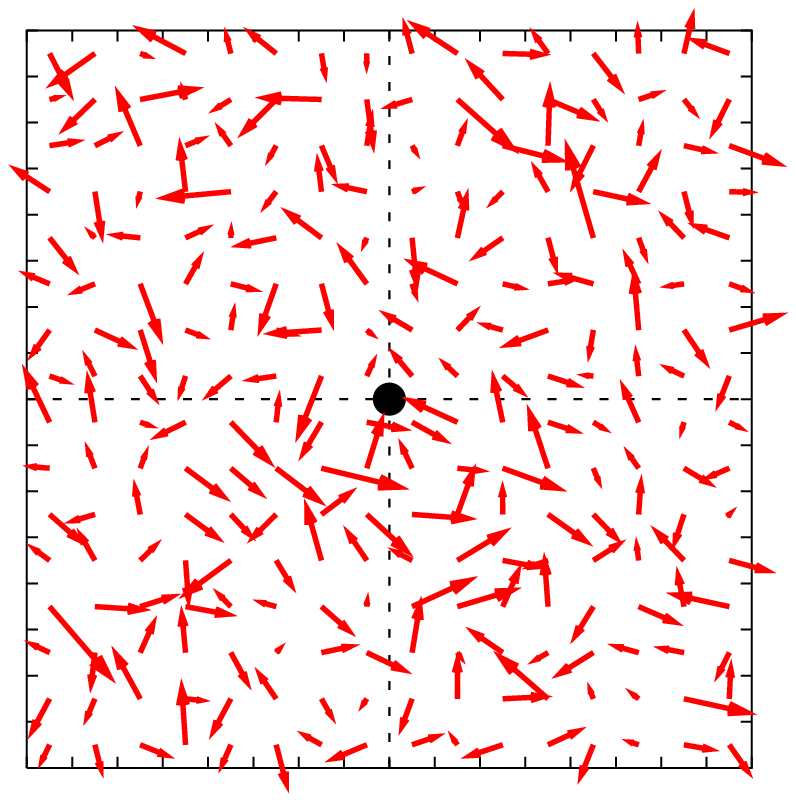}
\includegraphics[width=45mm]{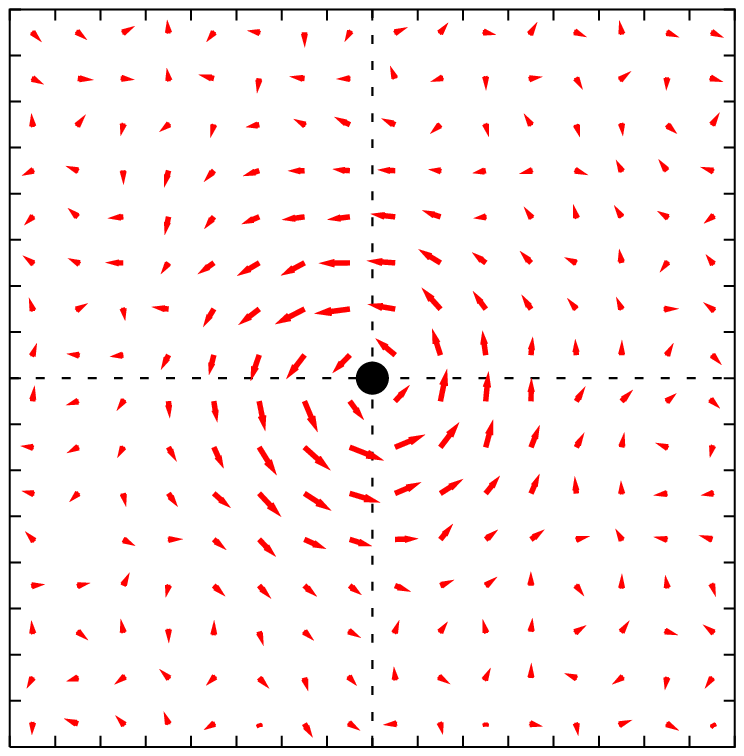}
\includegraphics[width=45mm]{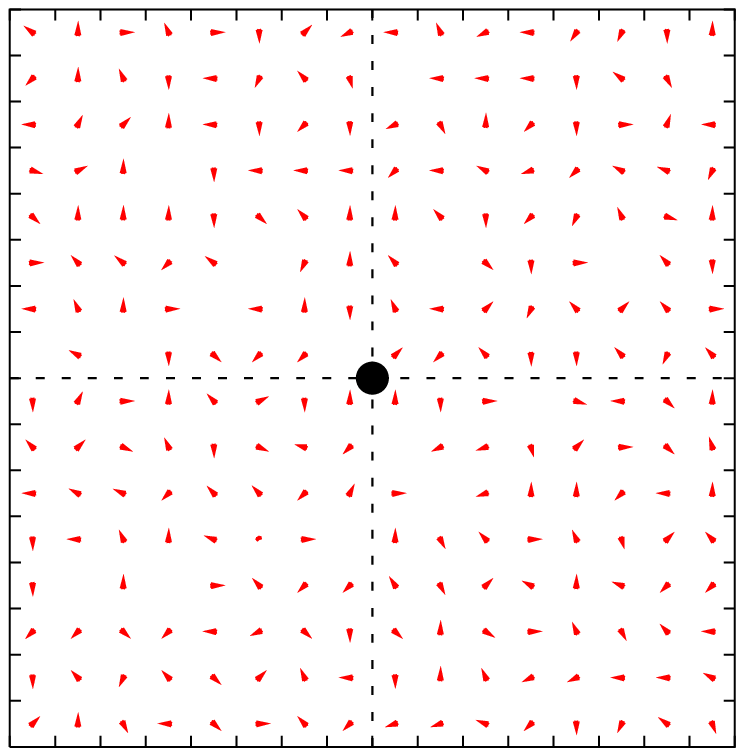}\\
\includegraphics[width=45mm]{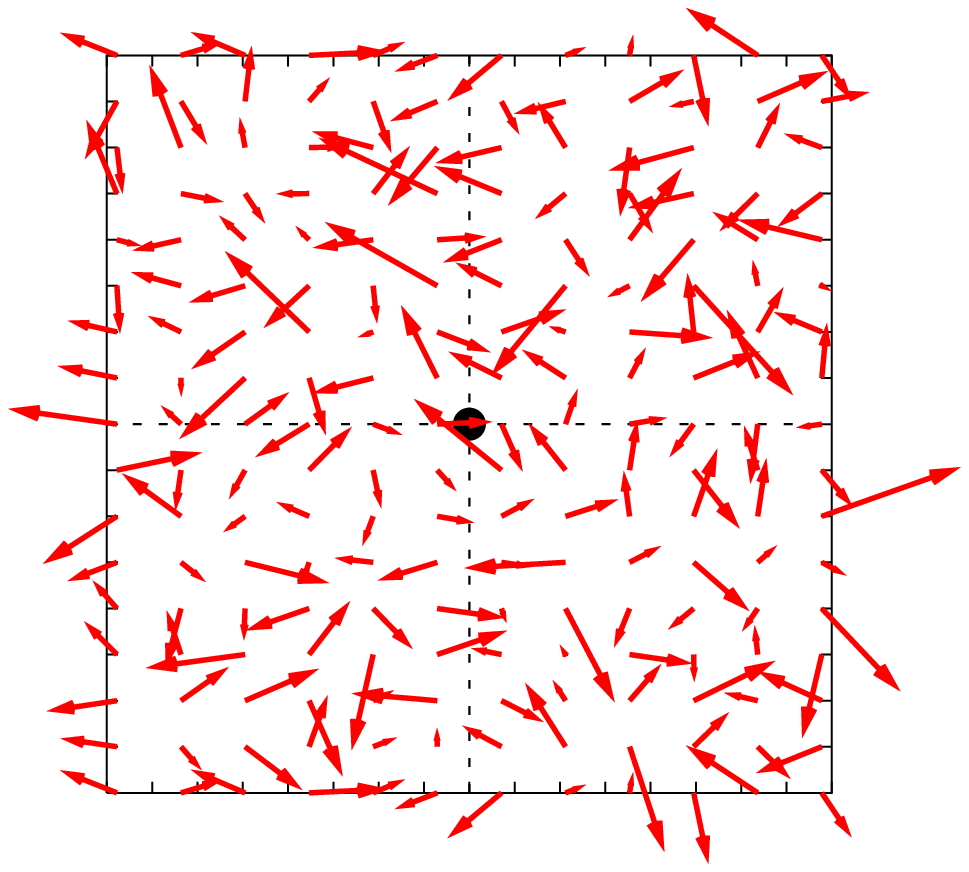}
\includegraphics[width=45mm]{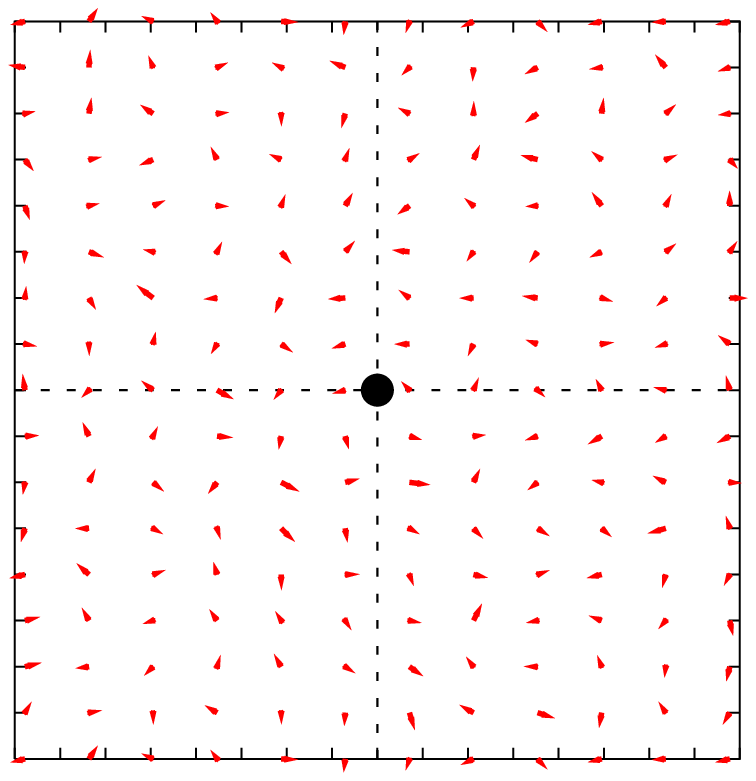}
\includegraphics[width=45mm]{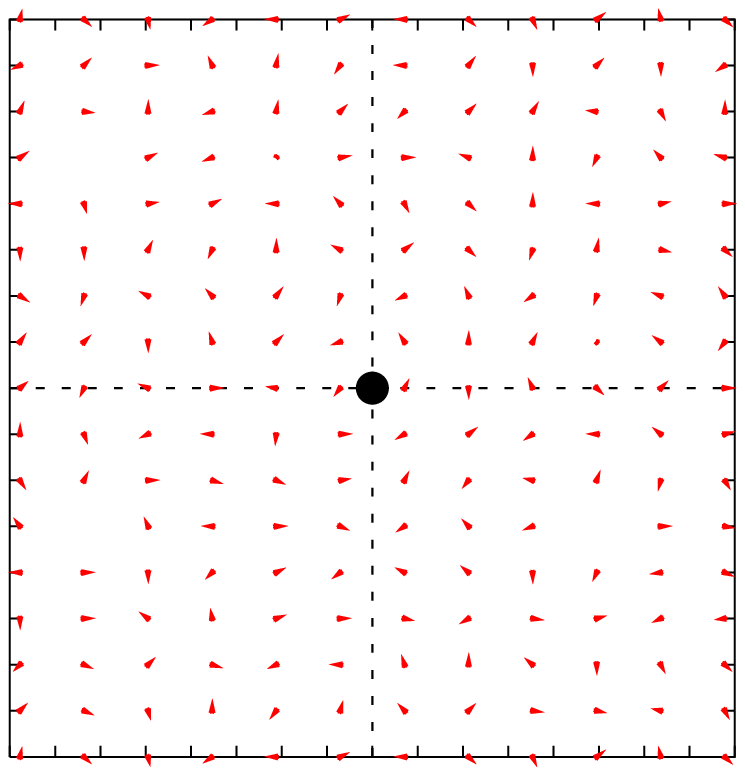}\\
\end{center}
\vspace{-15mm}
\caption{
The profile of the monopole current for the mesonic flux tube at $T/T_{c}=0.94$.
The sources and distances are as in Figure \ref{fig:Action_density_MESON}.
}
\vspace{-5mm}
\label{fig:Monopole_current_MESON}
\end{figure}

In Figure \ref{fig:Electric_field_MESON} we show the profile of the color-electric field in the plane including the $Q\bar{Q}$ axis, whereas the azimuthal component of the monopole current is plotted in Figure \ref{fig:Monopole_current_MESON}.
The figures suggest that the color-electric field is squeezed into a flux tube due to the magnetic supercurrent circulating around the $Q\bar{Q}$ axis and creating the solenoidal electric field.
The agreement with this picture has been found before in SU(2) gluodynamics and in both quenched and full QCD at zero temperature \cite{Koma_2002,Koma_2003,DIK_2003_ZERO,Ichie:2003}.
Figures \ref{fig:Electric_field_MESON} and \ref{fig:Monopole_current_MESON} present first evidence that this picture is also correct in full QCD at finite temperature \cite{DIK_2003}.
One can see from Figure \ref{fig:Electric_field_MESON} that for two smaller distances
the monopole part of the electric field has the form of a sourceless solenoidal field\cite{DIK_2003_ZERO}.
This field changes direction at some distance away from the $Q\bar{Q}$ axis and cancels in this region the Coulomb field created by the photon part.
As a result the total Abelian electric field has the form of a flux tube.
Comparison of the monopole part of the electric field for $R/r_{0}=0.91$ and $R/r_{0}=1.59$ reveals that the electric field becomes weaker with increasing distance.
It virtually disappears for $R/r_{0}=2.25$, thus providing clear evidence for the string breaking at $R/r_{0}=2.25$.
\begin{figure}[tpb]
\begin{center}
\includegraphics[width=150mm]{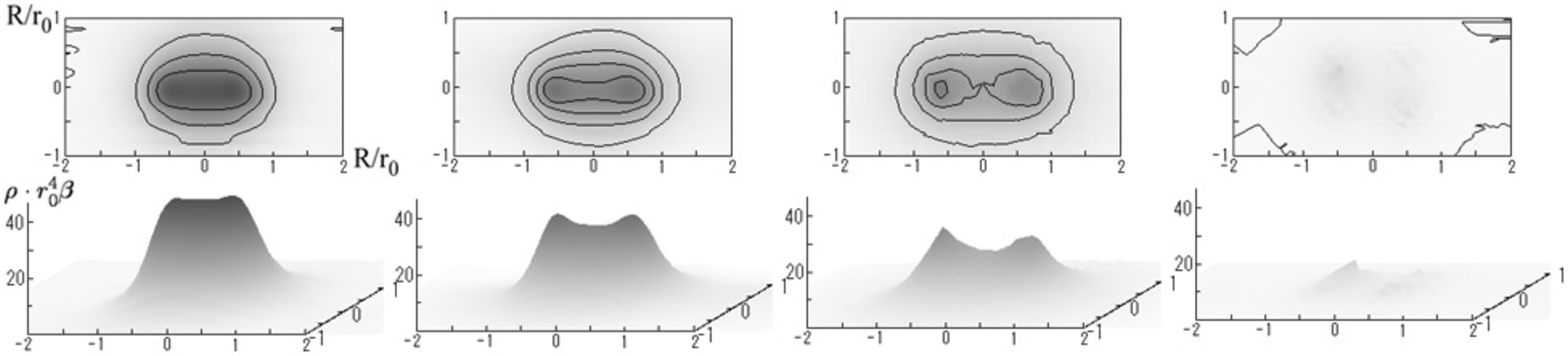}\\
\includegraphics[width=40mm]{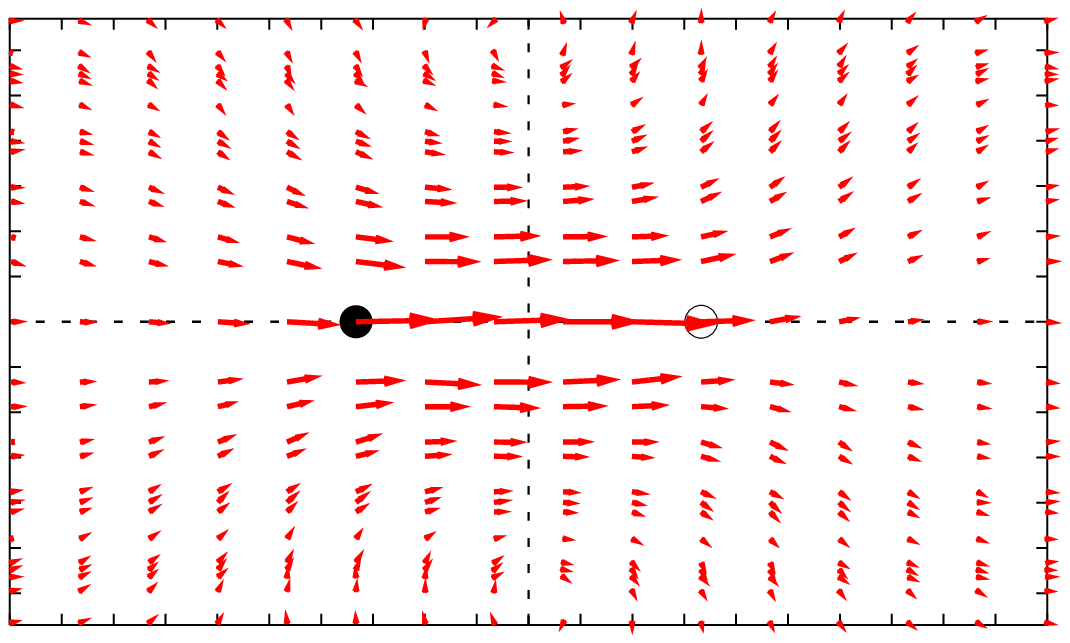}
\hspace{-5mm}\includegraphics[width=40mm]{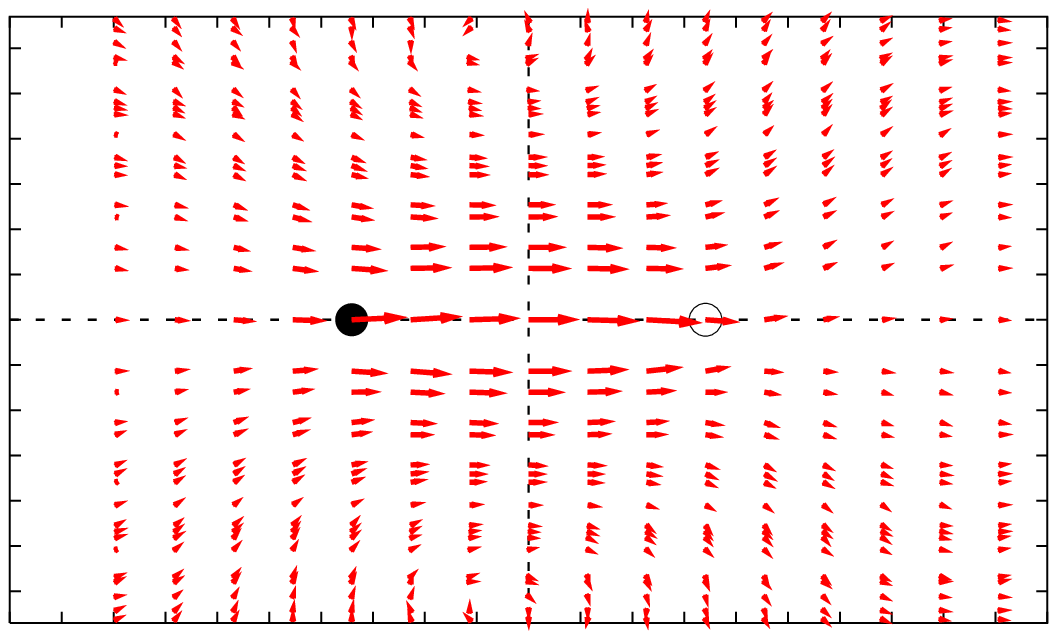}
\hspace{-5mm}\includegraphics[width=40mm]{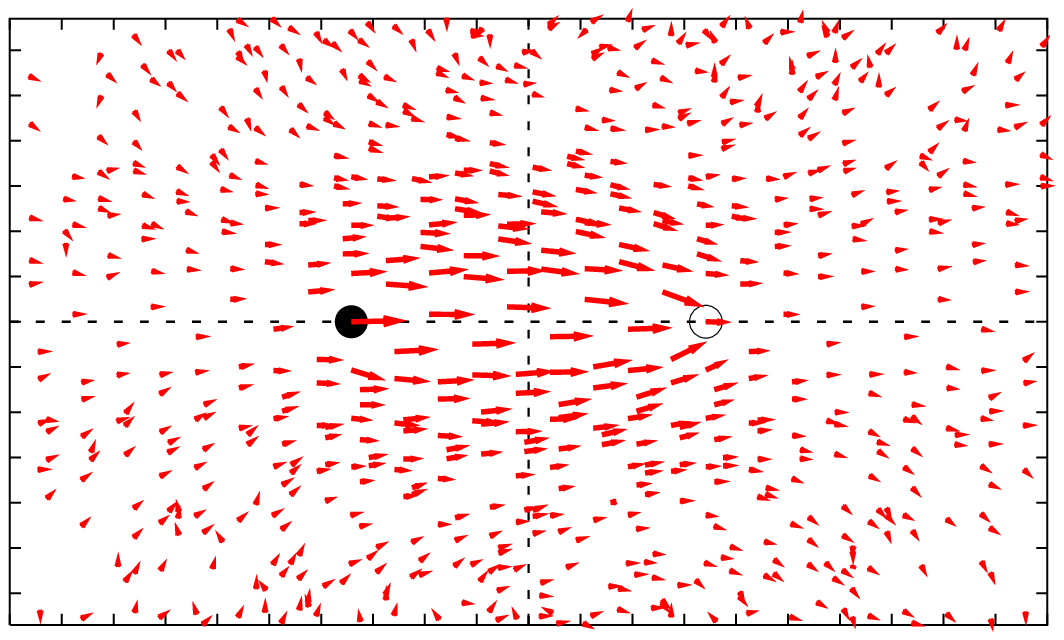}
\hspace{-5mm}\includegraphics[width=40mm]{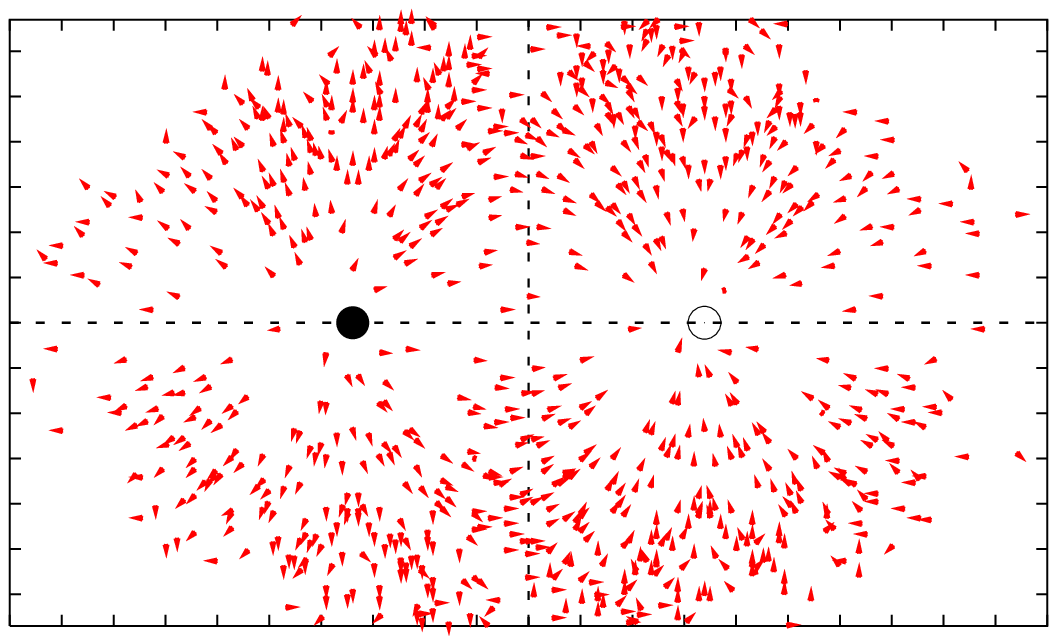}\\[-3mm]
\includegraphics[width=43mm]{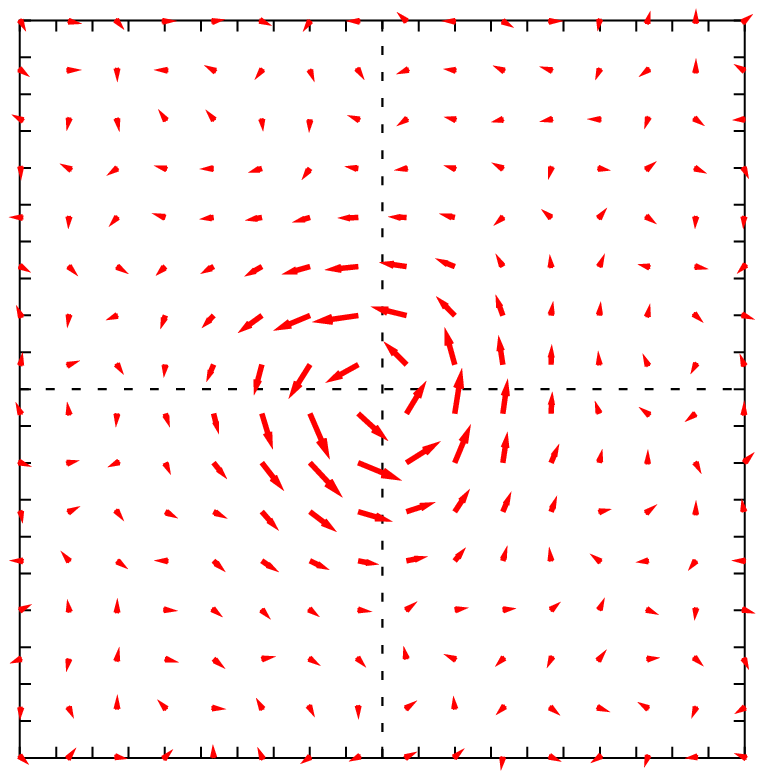}
\hspace{-8mm}\includegraphics[width=43mm]{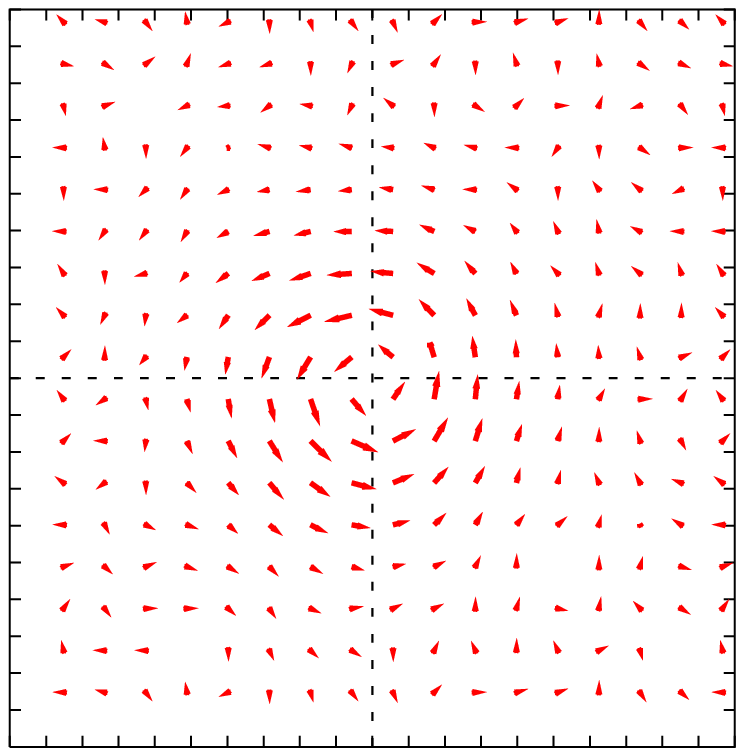}
\hspace{-8mm}\includegraphics[width=43mm]{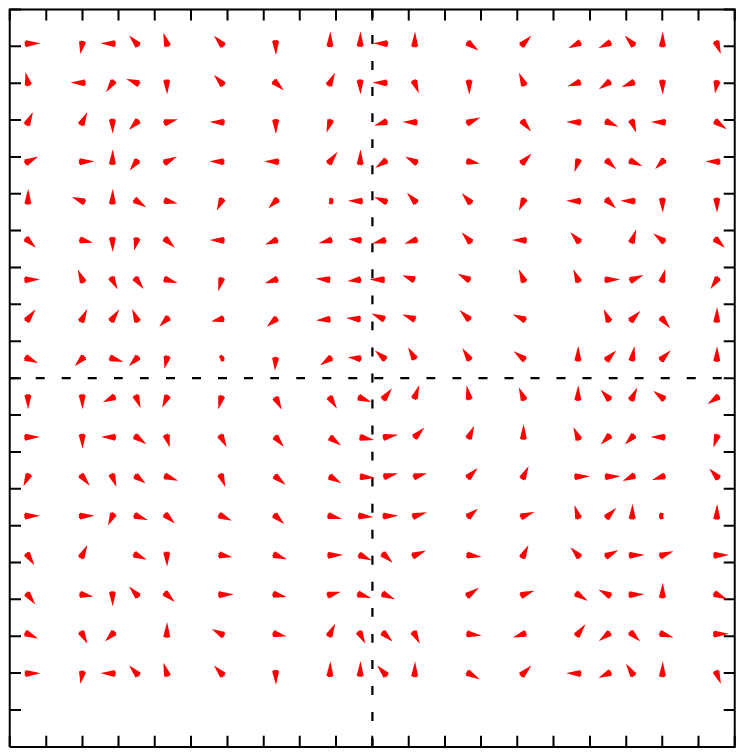}
\hspace{-8mm}\includegraphics[width=43mm]{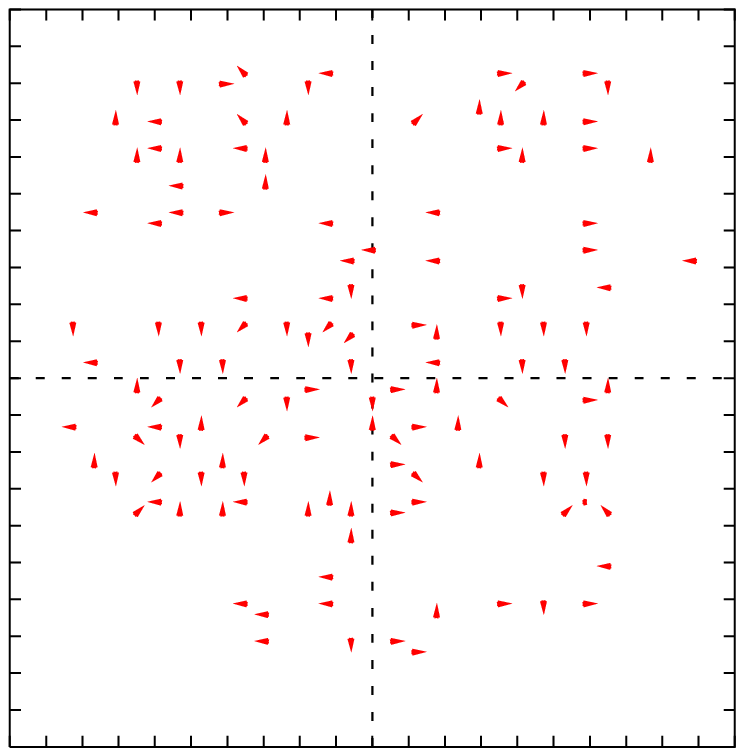}\\[-5mm]
\end{center}
\vspace{-15mm}
\caption{
The profile of the monopole part of the action density (top), the color-electric field (middle) and the monopole current (bottom) at (left to right) $T/T_{c}=0.80$, 0.94, 1.00 and 1.28 for $R/r_0\sim 1.36$.
}
\vspace{-5mm}
\label{fig:Phase_transition}
\end{figure}
String breaking at this distance can also be seen from the monopole part of the magnetic current, depicted in Figure \ref{fig:Monopole_current_MESON}.
Indeed the azimuthal monopole current disappears at this distance, while it is clearly seen at smaller distances.
It is known from studies of the SU(2) theory that the magnetic current calculated from monopole sources reproduces almost exactly that calculated with Abelian sources.
This can also be seen from the first row of Figure \ref{fig:Monopole_current_MESON}.
In the second row the Abelian part shows only random fluctuations which is due to stronger noise for this case in comparison with the monopole part.
For the photon part, as expected, we see only noise at all distances.

\subsection{Temperature dependence of the mesonic system}
~\\[-8mm]

It is also interesting to see how the flux tube of given length changes with the temperature.
We measured the same observable as in the previous subsection varying now the temperature, while keeping the $Q\bar{Q}$ distance $R$ approximately constant.

We show in Figure \ref{fig:Phase_transition} the profile of the monopole part only because the signal for it is much better than that for the Abelian part.
For $T/T_c=0.80$ and 0.94 we see the clear flux tube in all our observables.
We can see also that all observables decrease with increasing temperature.
This behavior is similar to the flux tube evolution with increasing distance discussed in the previous subsection.

At the critical temperature $T=T_c$, there seems to remain a touch of a flux tube as seen from the action density, although signals of the electric field and the monopole current are very weak.
Note that the existence of a weak flux tube at this temperature would be in agreement with the continuous character of the transition, which is expected to be a crossover.

In the deconfinement phase at $T/T_c=1.28$, no tube profile is observed.
This is consistent with the disappearance of the linear potential at $T>T_c$.

\section{Baryonic system}
\begin{figure}[tpb]
\begin{center}
\includegraphics[keepaspectratio=true,height=80mm]{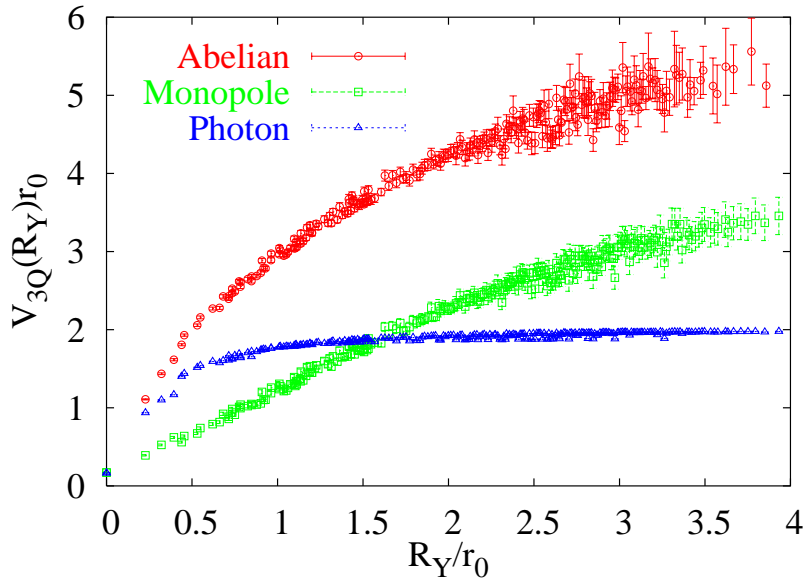}
\end{center}
\vspace{-12mm} \caption{The static three quark potential at $T/T_{c}=0.94.$}
\vspace{-5mm}
\label{fig:CR_QQQ_k1340.eps}
\end{figure}
\subsection{Static potential}
~\\[-8mm]

Now we discuss the system constructed from three static quarks. We measure the static three-quark potential defined as:
\begin{equation}
 V_{3Q}(
\overrightarrow{s_{1}},\overrightarrow{s_{2}},\overrightarrow{s_{3}},
T)= - \frac{1}{T}
\ln \left\langle \sum\frac{1}{3!}
|\varepsilon_{abc}|L^{a}(\overrightarrow{s_{1}}) L^{b}(\overrightarrow{s_{2}})
L^{c}(\overrightarrow{s_{3}}) \right\rangle\,. \label{eq:QQQ_potential}
\end{equation}
Since the baryonic flux tube has Y-shape at $T=0$~\cite{Ichie:2002mi,Ichie:2003,Kuzmenko:2000bq}, we use $r_Y$ defined by Eq.~(\ref{eq:R_min}) as a typical distance describing the three-quark system.

\begin{figure}[tpb]
\begin{center}
\includegraphics[width=120mm,clip=true]{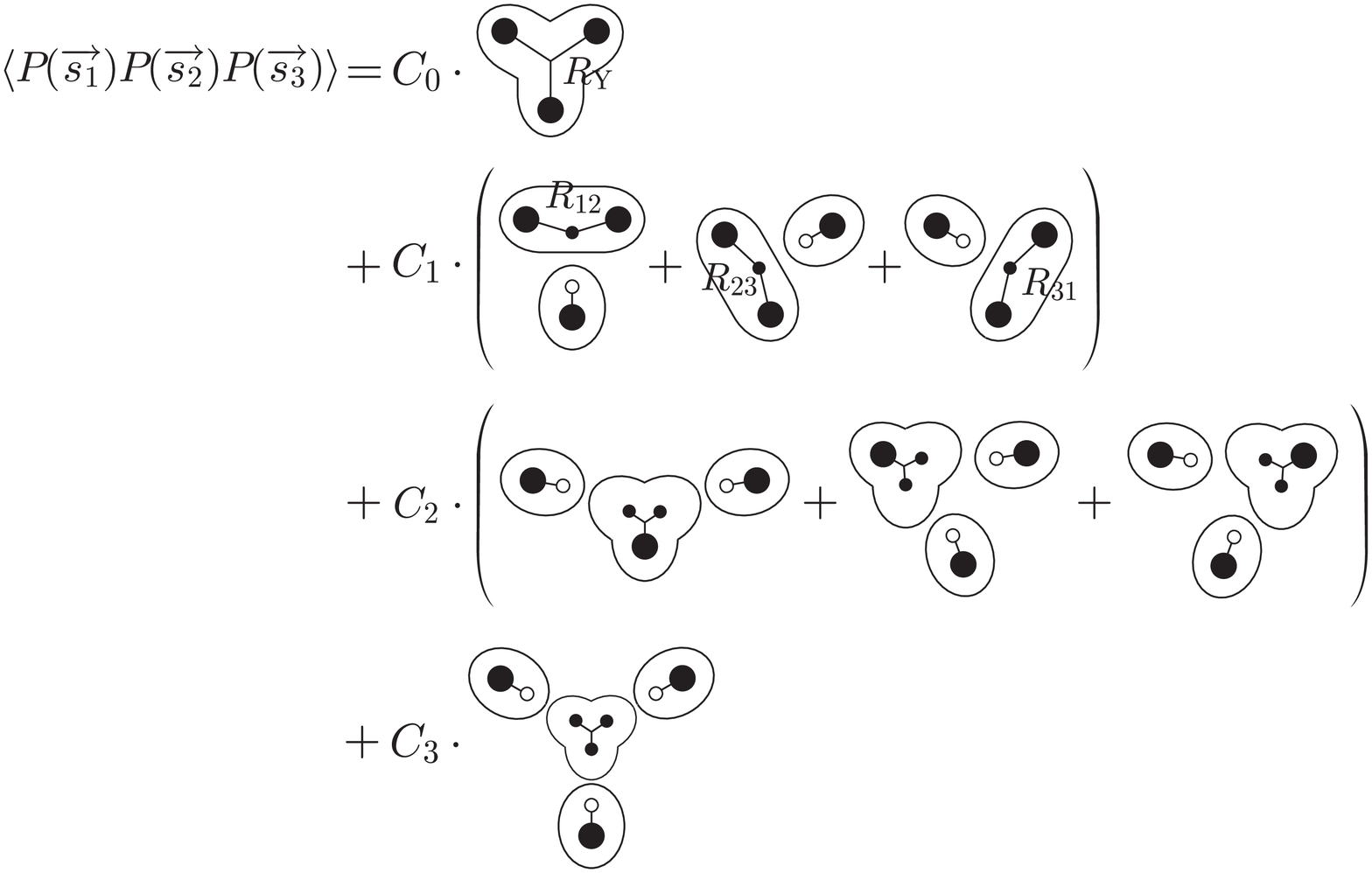}
\end{center}
\vspace{-12mm}
\caption{
Terms contributing to the correlation function of three Polyakov loops (\ref{eq:QQQ_potential}).
The large circles represent the static quarks $(Q)$, the small circles correspond to the dynamical quarks ($q$).
}
\label{fig:QQQ_breaking}
\end{figure}

Note that the Y-shape of the flux tube is also supported by results for the static potential in quenched QCD~\cite{Takahashi_2002}\footnote{However the data of
ref.~\cite{Alexandrou_2002} give preference to $\Delta$ type potential}.
As is seen from Figure~\ref{fig:CR_QQQ_k1340.eps} the static three quark potential has a tendency to flatten at large $r_{\rm Y}$.
At finite temperature not only the lowest energy state (baryon) but also
states of broken string, shown in Fig.~\ref{fig:QQQ_breaking}, may contribute to the potential~(\ref{eq:QQQ_potential}).

The breaking of the string occurs due to creation of pairs made from a light quark and a light anti-quark, $q\bar{q}$, from the vacuum.
Contrary to the mesonic case discussed in the previous section, breaking of the baryonic string may happen at seven different stages, depicted schematically in Figure~\ref{fig:QQQ_breaking}.

\begin{enumerate}
\item Process $QQQ+q\bar{q} \to QQq+Q\bar{q}$.\\
The most energetically favorable string breaking corresponds to creation of one virtual meson, $q\bar{q}$, from the vacuum.
The light quark from this pair and two heavy quarks form a baryon system, $QQq$.
The chromoelectric string inside the $QQq$--baryon is spanned between the two heavy quarks. The light anti-quark and the third heavy quark form a heavy--light, $Q\bar{q}$, meson state.
There are three possible ways to break the Y-string by one light meson, they are shown in the second row of Figure~\ref{fig:QQQ_breaking}.

\item Process $QQQ+ 2q\bar{q} \to Qqq+2Q\bar{q}$.\\
Another possibility is the creation of two light meson states, $2q\bar{q}$, from the vacuum.
Then the string becomes completely broken and the breaking produces two heavy--light mesons, $Q\bar{q}$ and one baryon made from the heavy quark and the two light quarks, $Qqq$. There are also three possible combinations of the final states depicted in the third row of Figure~\ref{fig:QQQ_breaking}.

\item Process $QQQ+ 3q\bar{q} \to qqq+3Q\bar{q}$.\\
Finally, three light mesons can be created from the vacuum producing three heavy--light mesons~$Q\bar{q}$ and one $qqq$--baryon made from the light quarks (the bottom of Figure~\ref{fig:QQQ_breaking}).
\end{enumerate}

All these states must be taken into account in the correlator of the three static quarks.
Thus instead of the two-exponential formula~(\ref{eq:two:exp}) we will have the five-exponential ansatz which includes also the original unbroken string state (the first term below):
\begin{eqnarray}
\langle L(\overrightarrow{s_1})L(\overrightarrow{s_2})L(\overrightarrow{s_3}) \rangle
&=& e^{- V_0/T} \Bigl\{
e^{- \sigma_{QQQ} R_{\rm Y}/T}\nonumber\\
&+& e^{- m_{Q\bar{q}}/T - m_{QQq}/T}\, \left[
e^{- \sigma_{QQq} R_{12}/T}+e^{- \sigma_{QQq} R_{23}/T}+e^{- \sigma_{QQq} R_{31}/T}\right]\nonumber\\
&+& 3 \, e^{- 2 \, (m_{Q\bar{q}} +m_{Qqq})/T}
\label{eq:three:exp}\\
&+& e^{- 3 \, (m_{Q\bar{q}}+ m_{qqq})/T} \nonumber\Bigl\}\,.
\end{eqnarray}
This simplified formula takes into account only the linear potentials between the heavy quarks as well as the broken string states, disregarding other possible interactions between the quarks.
It is known \cite{mon:st,Bali:1996dm,DIK_2003_ZERO} that both in quenched and unquenched QCD the monopole part of the mesonic potential is linear at large as well as at small distances.
In the study of the mesonic potential in full QCD at finite temperature \cite{DIK_2003} we found again that the monopole part of the potential can be fitted well by Eq.(\ref{eq:two:exp}) which includes only the linear potential in $V_{string}(r,T)$.
Thus, we may expect that the correlator (\ref{eq:QQQ_potential}) of the monopole Polyakov loops~(\ref{eq:PL_op_Mo}) can be described by the simple form of Eq.~(\ref{eq:three:exp}).

In Eq.~(\ref{eq:three:exp}) we explicitly used two different string tensions, one for the
tension of the Y-shaped unbroken string state (the first term in the {\it r.h.s.} of (\ref{eq:three:exp})) and another for the string tension of the $QQq$ baryon.
On the classical level there is no difference between strings inside $QQQ$ and $QQq$ states.
The only difference is in the shape of strings : the $QQQ$ state has Y-shape string while in the $QQq$ state the chromoelectric string between heavy quarks forming an approximated straight line.
In the $QQq$ state the sea quark, $q$, should be located somewhere inside the string.
However, on the quantum level the sea quark may interact with the string and the string tension may, in general, get an additional correction from the quantum motion of the sea quark.

In Eq.(\ref{eq:three:exp}) the dependence on the self-energy $V_0$ is factorized similarly to the mesonic case~(\ref{eq:two:exp}).
The expression in the curly brackets contains terms depending on the constituent masses of light quarks in the $Q\bar q$, $QQq$, $Qqq$ and $qqq$ states ($m_{QQq}$, $m_{Q\bar{q}}$, $m_{Qqq}$ and $m_{qqq}$, respectively).
The potential~(\ref{eq:three:exp}) depends not only on $R_{\rm Y}$, but also on $R_{ij}$ and this is the reason of the broad distribution of points in Fig.~\ref{fig:CR_QQQ_k1340.eps}.
Note that the two completely broken string states, $Qqq+2Q\bar{q}$ and $qqq+3Q\bar{q}$ (corresponding to the last two rows in Figure~\ref{fig:QQQ_breaking}) provide distance--independent contributions to the potential~(\ref{eq:three:exp}).
This independence exists only for the monopole part of the potential, which neglects the Coulomb interactions between quarks, and, consequently, van der Vaals forces between the mesons and the baryons.
Thus, for the non-Abelian potential or for the corresponding Abelian part of it the contribution of the $Qqq+2Q\bar{q}$ and $qqq+3Q\bar{q}$ states to the potential should be dependent on the distances between the quarks.

\begin{figure}[tpb]
\begin{center}
\centerline{\includegraphics[width=80mm,clip=true]{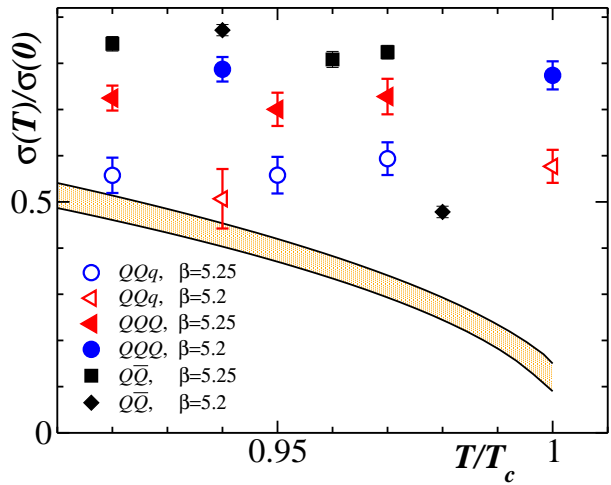}}
\end{center}
\vspace{-18mm}
\caption{
The monopole contribution to tensions of chromoelectric strings inside the $QQQ$ and $QQq$ states for $\beta=5.2$ and $\beta=5.25$.
We show for comparison the tension of the string obtained in the meson state ($Q\bar Q$)~\cite{DIK_2003}, and the shaded area shows the quenched value of the $Q\bar Q$ string tension~\cite{Kaczmarek:2000mm}.
}
\vspace{-5mm}
\label{fig:QQQ:string:tension}
\end{figure}
%
We have analyzed the data for the three quark correlation function following Ref.~\cite{Takahashi_2002}, but with the ansatz~(\ref{eq:three:exp}) which takes into account the
breaking of the string.
The results for both string tensions are presented in Figure~\ref{fig:QQQ:string:tension}.
We show for comparison the tension of the string obtained in the mesonic case in $N_f=2$ QCD ~Ref.~\cite{DIK_2003} and in the corresponding quenched value~\cite{Kaczmarek:2000mm}.
One can see that string tensions are almost independent of the temperature in the region close to the phase transition.
Moreover, the string tension of the $QQQ$ system is slightly higher than the string tension of the $QQq$ state.
Thus, we conclude that the sea quark lowers the energy of the string.

Another interesting result presented in
Figure~\ref{fig:QQQ:string:tension} is that the tension of the
unbroken string in the baryon, $\sigma_{QQQ}$, is slightly
lower than the tension of the string in meson,
$\sigma_{Q\bar{Q}}$.

\begin{figure}[tpb]
\begin{center}
\centerline{\includegraphics[width=80mm,clip=true]{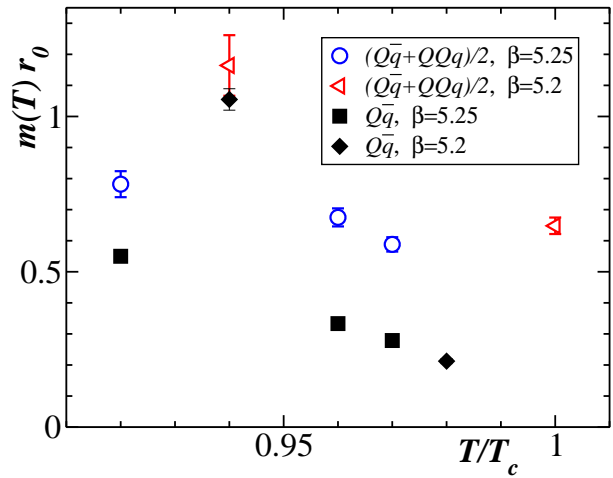}}
\end{center}
\vspace{-18mm}
\caption{
The same as in Figure~\ref{fig:QQQ:string:tension} but for the average of the effective quark masses in the $Q\bar q$ meson and the $QQq$ baryon, $(m_{Q\bar{q}}+m_{QQq})/2$.
We also show the effective quark masse of the $Q\bar q$ meson obtained from two-point Polyakov loop correlators in Ref.~\cite{DIK_2003}.
}
\vspace{-5mm}
\label{fig:QQQ:masses}
\end{figure}

Using the fitting function~(\ref{eq:three:exp}) we obtain the sum of the constituent masses of the light quarks in the $Q\bar{q}$ and $QQq$ states.
As one can see from the form of the fitting function~(\ref{eq:three:exp}) we can not get the values of these masses (as well as the other mass parameters) separately using the monopole contribution to the correlation function of the three quarks.
In Figure~\ref{fig:QQQ:masses} we present the averaged constituent quark mass, $(m_{Q\bar{q}}+m_{QQq})/2$, which shows a tendency to drop with increasing temperature.
We also depict in this Figure the constituent quark mass in the $Q\bar q$ meson which was obtained from two-point Polyakov loop correlators in Ref.~\cite{DIK_2003}.
Our data shows that the constituent quark mass in the $QQq$--baryon is higher than in the mass in the heavy--light meson, $Q\bar{q}$.
%

\subsection{Profiles of the baryonic system}
~\\[-8mm]

\begin{figure}[tpb]
\begin{center}
\includegraphics[width=150mm]{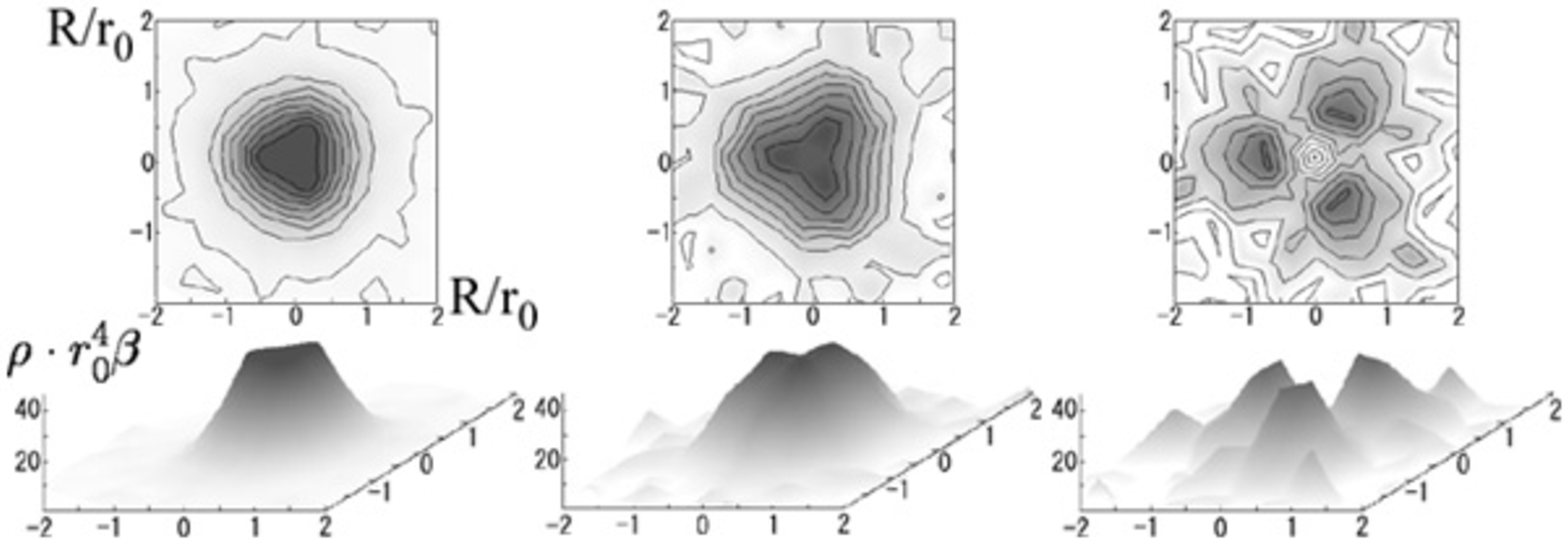}\\
\hspace{3mm}\includegraphics[width=40mm]{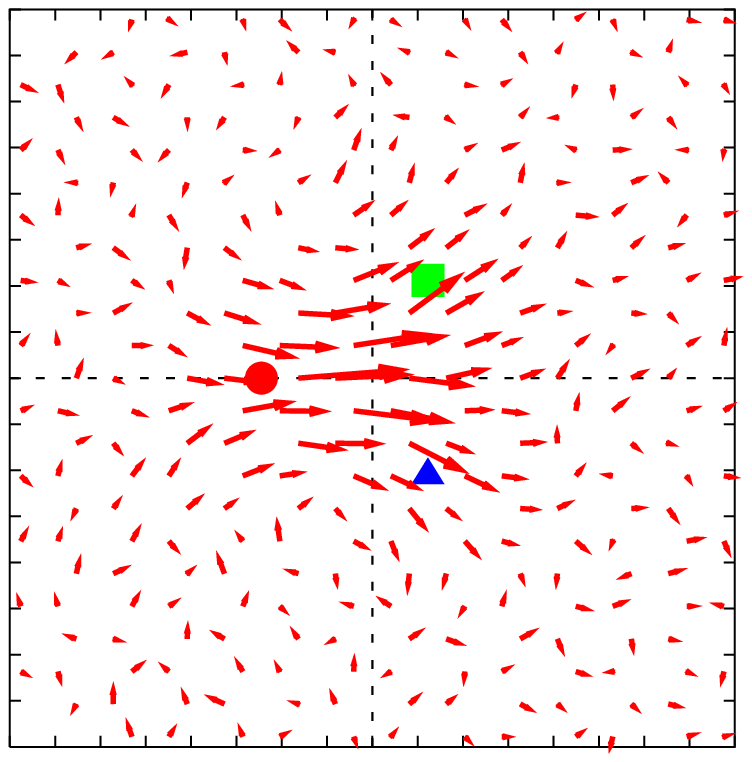}
\hspace{6mm}\includegraphics[width=40mm]{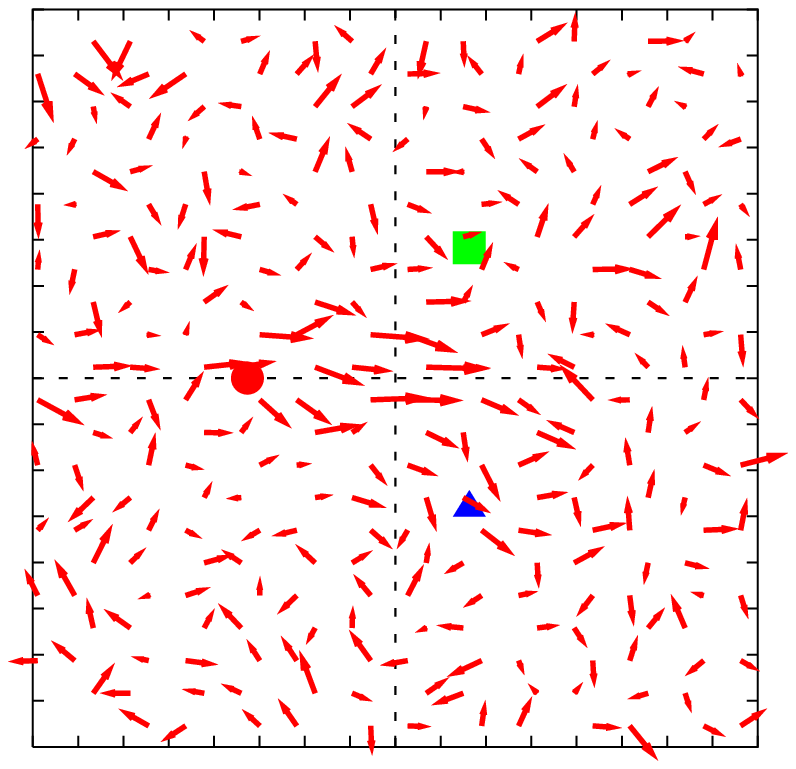}
\hspace{6mm}\includegraphics[width=40mm]{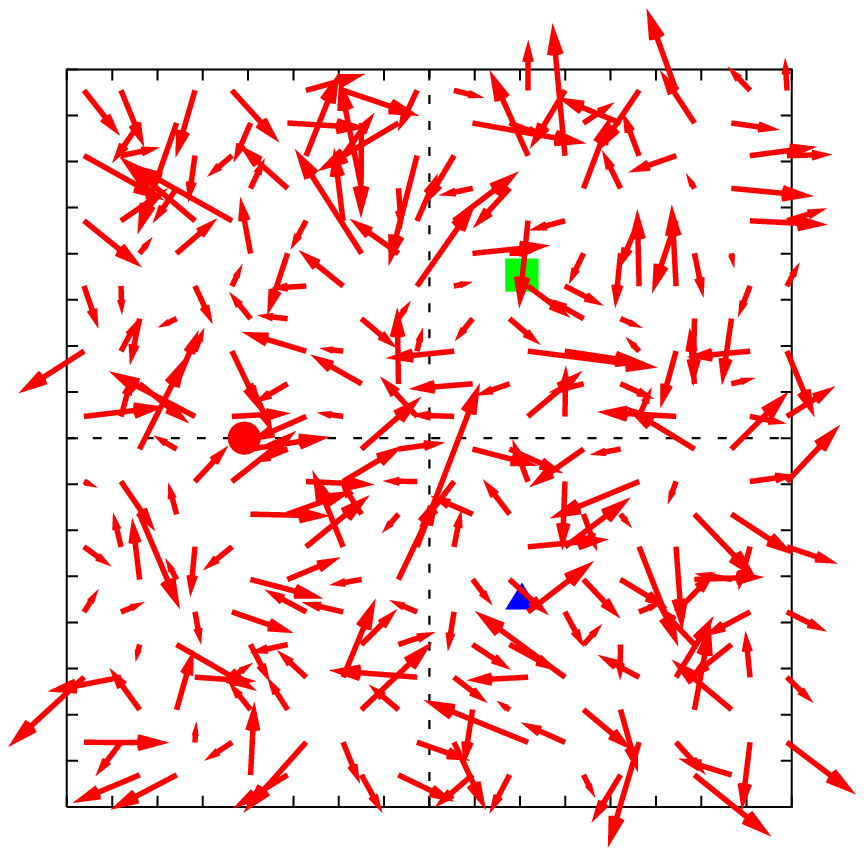}
\end{center}
\vspace{-13mm}
\caption{
The action density (top) and the color-electric field (bottom) of the baryonic system.
The sources are made of the monopole part of Polyakov loops.
The distances are $R_{\rm Y}/r_{0}=1.96$(left), 2.61(center) and  3.26(right), and the temperature is fixed to $T/T_{c}=0.80\ ~(\kappa=0.1330)$.
}
\vspace{-5mm}
\label{fig:BARYON_confinement_phase}
\end{figure}

In Fig.\ref{fig:BARYON_confinement_phase} we show profiles of the action density and the color-electric field in the baryonic system.
For the static quark sources we use the monopole Polyakov loops.
For $R_{\rm Y}/r_{0}=1.96$, 2.61 the linear part of the static potential exists and the profile of the action density has Y-shape.
The electric field is also squeezed into Y-shape.
However, the noise is much bigger than that for the $T=0$ case~\cite{Ichie:2002mi,Ichie:2003}.

We are using the correlator of three Polyakov loops to describe the baryonic system at finite temperature. 
Thus there is no a Y-shape junction in the operator itself, contrary to the case of the Wilson loop used at zero temperature~\cite{Ichie:2002mi,Ichie:2003}.
Nevertheless we clearly see that the flux distribution has a Y-shape geometry with a junction, similar to the zero temperature case \cite{Ichie:2002mi,Ichie:2003}.

For large enough $R_{\rm Y}$ string breaking is expected.
We see this effect in the action density in Fig.\ref{fig:BARYON_confinement_phase} at $R_Y/r_0=3.26$.
The color-electric field is too noisy though to observe string breaking.
\begin{figure}[tpb]
\begin{center}
\includegraphics[width=90mm]{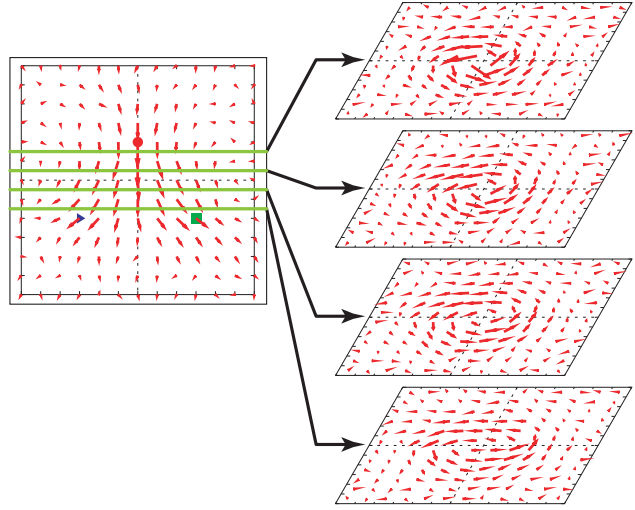}
\end{center}
\vspace{-15mm} \caption{ The profile of the color-electric field (left) and the
monopole current (right) at $T/T_{c}=0.87$ in a three-quark system. }
\vspace{-7mm} \label{fig:BARYON_structure}
\end{figure}
\begin{figure}[tpb]
\begin{center}
\includegraphics[width=155mm]{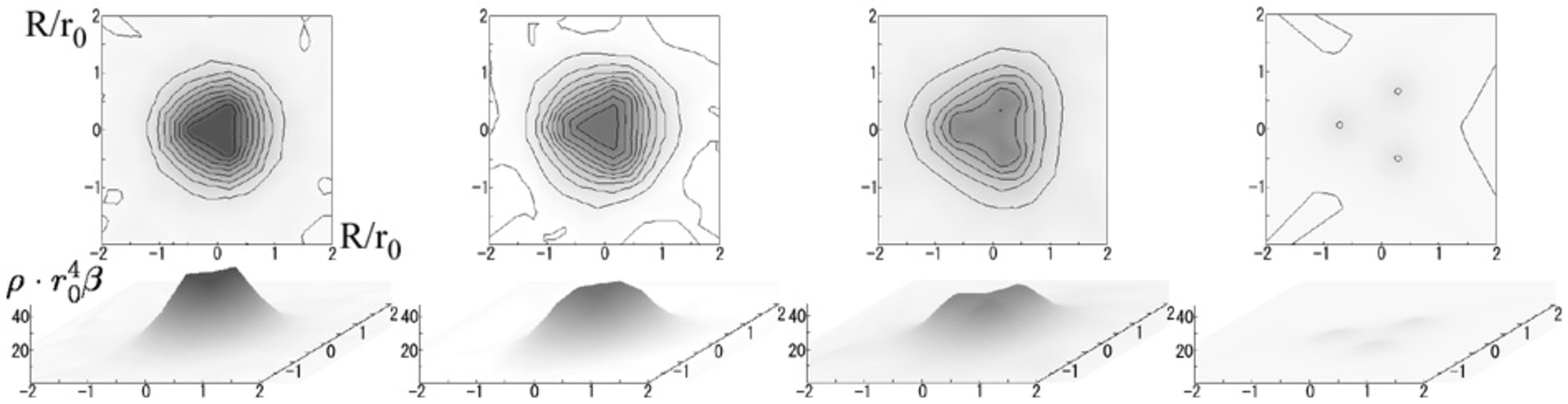}\\
\includegraphics[width=45mm]{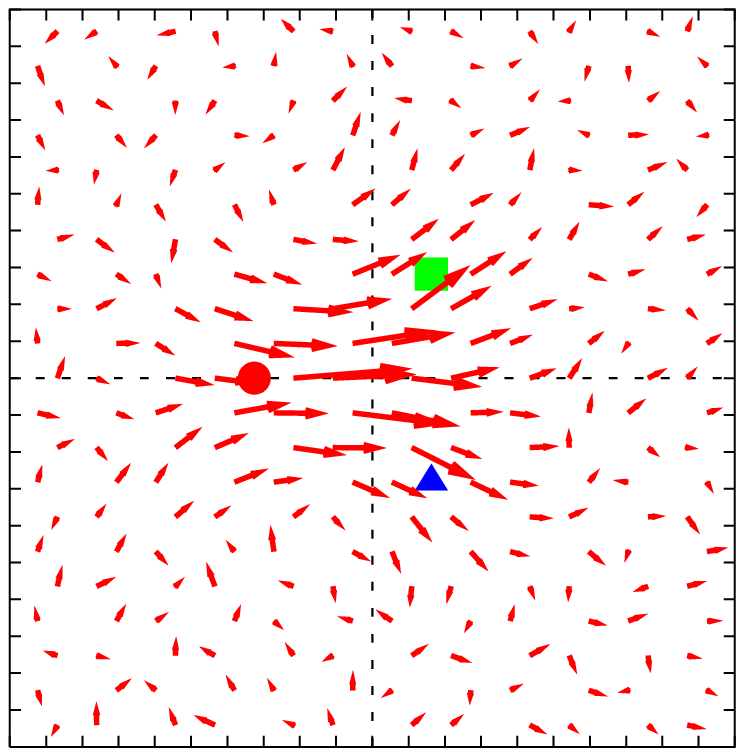}
\hspace{-9mm}\includegraphics[width=45mm]{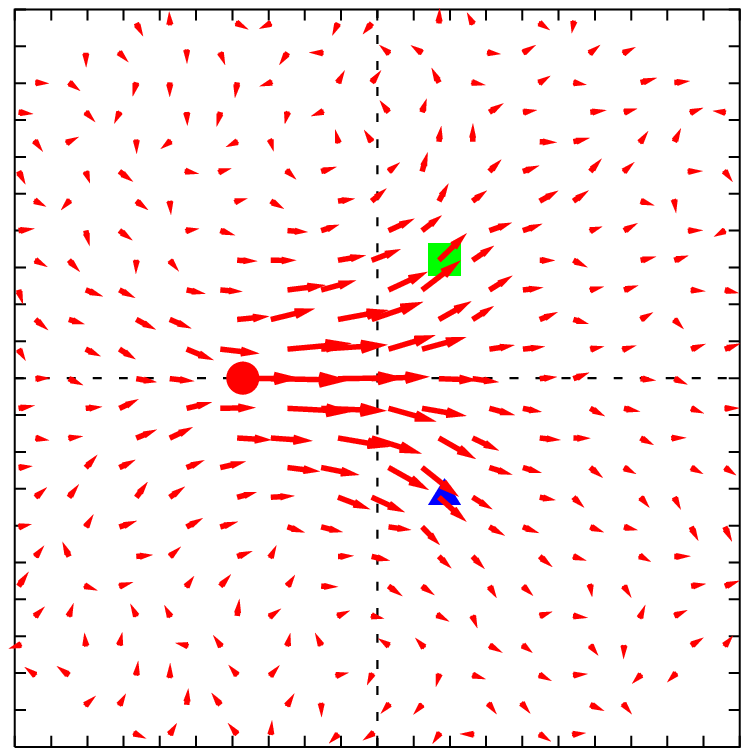}
\hspace{-9mm}\includegraphics[width=45mm]{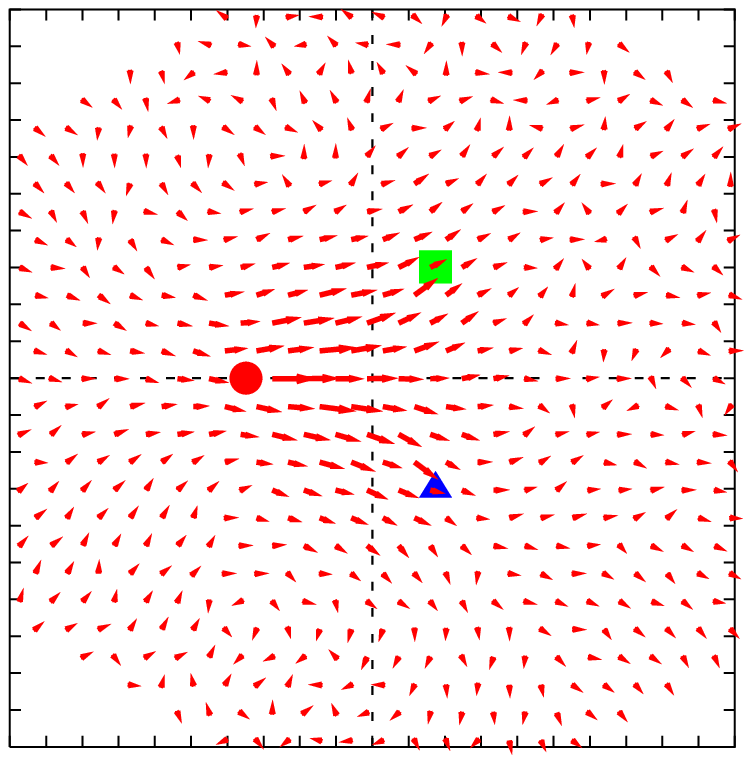}
\hspace{-9mm}\includegraphics[width=45mm]{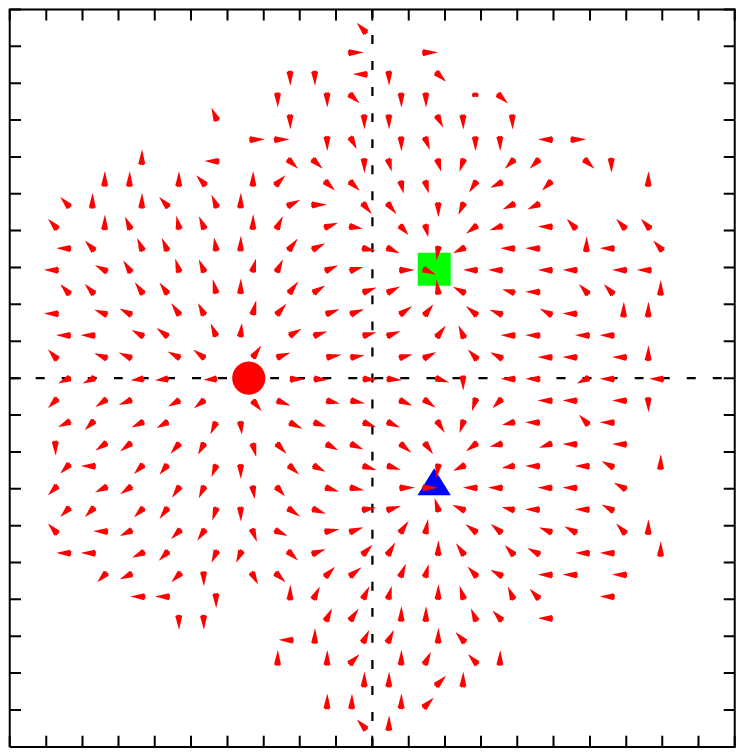}\\
\end{center}
\vspace{-15mm}
\caption{
The profile of the monopole part in the baryonic system at (left to right) $T/T_{c}=0.80$, 0.94, 1.00 and 1.28.
These figures correspond to physical distance $R_{{\rm Y}}/r_0\sim2.0$.
The top row shows the action density, the bottom row shows the color-electric field.
}
\vspace{-5mm}
\label{fig:BARYON_DCP_transition}
\end{figure}
In Fig.\ref{fig:BARYON_structure} we show the color-electric field and the monopole current distributions for $T/T_{c}=0.80$ and $R_{\rm Y}/r_{0}=1.96$.
The electric field is squeezed into a Y-shape flux tube as we already discussed.
One can see circulating monopole currents around the color-electric field in each slice.
In the plane where the color-electric field is subdivided into two parts, the circulating monopole current is not a simple circle anymore.
This indicates the possibility for the formation of two circulating currents if the distance between quarks is large enough.

In Fig.\ref{fig:BARYON_DCP_transition} we plot the profile of the baryonic system at various temperatures for $R_{{\rm Y}}/r_0\sim 2.0$, monopole Polyakov loops represent static quarks.
The behavior of the flux tube profile is similar to the meson case.
When the temperature increases, the density of the action and the electric field decrease and almost disappear in the deconfinement region.

\section{Conclusions}
We have studied the Abelian profiles of the action density, the electric field and the monopole current in the mesonic and the baryonic systems below and above the finite temperature transition in $N_f=2$ full QCD on $N_t=8$ lattices.
For the source we used Abelian, monopole and photon Polyakov loop operators.
In the mesonic case, we see a clear flux tube profile for some, not too large, quark-anti-quark distances $R$.
However, even in the confinement phase, flux tube breaking is observed for large enough $R$, in agreement with expectations.
We have also studied the temperature dependence of the mesonic flux tube profile keeping the distance $R$ fixed at about 0.7 fm.
We found that the value of the flux density decreases smoothly as the temperature approaches $T_c$.
In the deconfinement phase, the flux tube disappears.
But at the critical temperature, $T_c$, the flux tube still exists.

In the baryonic case, we have calculated the static three-quark potential in finite temperature full QCD.
However, our statistics is not good enough to study the baryonic string breaking in detail. Nevertheless the flattening of the baryonic potential is actually observed for large $R_Y$.
The flux tube profile and the string breaking phenomenon are consistent with the meson case.
It is interesting that we could observe the Y-shape profile without introducing any junction in the baryon creation operator.

We are performing now the calculations on $24^3\times 10$ lattice to check for finite volume effects.
~\\[-5mm]
\section*{Acknowledgements}
This work is supported by the SR8000 Supercomputer Project of High Energy
Accelerator Research Organization (KEK). A part of numerical measurements has
been done using NEC SX-5 at RCNP of Osaka University. T.S. is partially
supported by JSPS Grant-in-Aid for Scientific  Research on Priority Areas
No.13135210 and (B) No.15340073. The Moscow group is partially supported by
RFBR  grants 02-02-17308, 01-02-17456, DFG-RFBR RUS 113 1739/0, RFBR-DFG
03-02-04016, grants INTAS--00-00111 DFG-RFBR 436 RUS 113/739/0, and CRDF awards
RPI-2364-MO-02 and MO-011-0.
\vspace{-1mm}


\begin{thebibliography}{99}
\newcommand{\plb}[1]{Phys. Lett. {\bf B#1}\ }
\bibitem{Aoki:1998sb}
S.~Aoki {\it et al.}  [CP-PACS Collaboration],
Nucl.\ Phys.\ Proc.\ Suppl.\  {\bf 73} (1999) 216.
\bibitem{Bolder:2000un}
B.~Bolder {\it et al.},
Phys.\ Rev.\ D {\bf 63} (2001) 074504
\bibitem{DIK_2003_ZERO}
DIK Collaboration (V.G. Bornyakov et al.),
hep-lat/0310011;
V.~Bornyakov {\it et al.},
Nucl.\ Phys.\ Proc.\ Suppl.\  {\bf 106} (2002) 634.
\bibitem{Bernard:2001tz}
C.~W.~Bernard {\it et al.},
Phys.\ Rev.\ D {\bf 64} (2001) 074509.
\bibitem{Bali_1994}
G.~S.~Bali, K.~Schilling and C.~Schlichter,
Phys.\ Rev.\ D {\bf 51} (1995) 5165;~
R.~W.~Haymaker, V.~Singh, Y.~C.~Peng and J.~Wosiek,
Phys.\ Rev.\ D {\bf 53} (1996) 389.
\bibitem{Ichie:2002mi}
H.~Ichie, V.~Bornyakov, T.~Streuer and G.~Schierholz,
Nucl.\ Phys.\ Proc.\ Suppl.\  {\bf 119} (2003) 751;
Nucl.\ Phys.\ A {\bf 721} (2003) 899.
\bibitem{Ichie:2003}
V. G. ~Bornyakov {\it et al.} [DIK Collaboration], [arXiv:hep-lat/0401026].
\bibitem{Takahashi_2002}
T. T. Takahashi, H. Suganuma, Y. Nemoto, H. Matsufuru,
Phys. Rev. D\textbf{65} (2002) 114509.
\bibitem{Kamizawa:1992hb}
S.~Kamizawa, Y.~Matsubara, H.~Shiba and T.~Suzuki,
Nucl.\ Phys.\ B {\bf 389} (1993) 563.
\bibitem{Chernodub:1998ie}
M.~N.~Chernodub and D.~A.~Komarov,
JETP Lett.\  {\bf 68} (1998) 117.
\bibitem{Kuzmenko:2000bq}
D.~S.~Kuzmenko and Y.~A.~Simonov,
Phys.\ Lett.\ B {\bf 494} (2000) 81.
\bibitem{Koma_2001_02}
Y. Koma, E.-M. Ilgenfritz, T. Suzuki and H. Toki
Phys. Rev. D64 (2001) 014015
\bibitem{DeTar:1998qa}
C.~DeTar, O.~Kaczmarek, F.~Karsch and E.~Laermann,
Phys.\ Rev.\ D {\bf 59} (1999) 031501.
\bibitem{DIK_2003}
V. G. ~Bornyakov {\it et al.} [DIK Collaboration], [arXiv:hep-lat/0401014],
V.~Bornyakov {\it et al.},
Nucl.\ Phys.\ Proc.\ Suppl.\  {\bf 119} (2003) 703.
\bibitem{Booth:2001qp}
S.~Booth {\it et al.}  [QCDSF-UKQCD collaboration],
Phys.\ Lett.\ B {\bf 519} (2001) 229.
\bibitem{Bali:1996dm}
G. S. ~Bali, V. G. ~Bornyakov, M. ~Muller-Preussker and K. ~Schilling,
Phys. Rev.  D {\bf 54} (1996) 2863
\bibitem{MaA}
A. S. ~Kronfeld,  M. L. ~Laursen,  G. ~Schierholz and U. J. ~Wiese,
Phys.\ Lett. {\bf B198}, 516 (1987).
\bibitem{ploop} T. ~Suzuki et al., \plb{347} (1995) 375;
[Erratum-ibid.\ B {\bf 351} (1995) 603].
\bibitem{Smit_1991} J. ~Smit and A. van der ~Sijs,
Nucl. Phys. \textbf{B355} (1991) 603.
\bibitem{Suzuki_1995} T. ~Suzuki, S. ~Ilyar, Y. ~Matsubara, T. ~Okude and K. ~Yotsuji,
Phys. Lett. \textbf{B347} (1995) 375;
Erratum: ibid. \textbf{B351} (1995) 603.
\bibitem{Gao:kg}
M.~Gao,
Phys.\ Rev.\ D {\bf 40} (1989) 2708; see also: Ph. de ~Forcrand, G. ~Schierholz,
H. ~Schneider and  M. ~Teper, Phys. Lett. B {\bf 160} (1985) 137.
\bibitem{Satz:2001kf}
H.~Satz,
arXiv:hep-ph/0111265.
\bibitem{Koma_2002}
Y. Koma, M. Koma, E.-M. Ilgenfritz, T. Suzuki, M.I. Polikarpov,
Phys. Rev. D\textbf{68} (2003) 094018,
\bibitem{Koma_2003}
Y. Koma, M. Koma, E.-M. Ilgenfritz, T. Suzuki,
Phys. Rev. D\textbf{68} (2003) 114504.
\bibitem{Alexandrou_2002}
C. Alexandrou, Ph. de Forcrand, A. Tsapalis,
Phys. Rev. D\textbf{66} (2002) 094503.
\bibitem{mon:st}
H.~Shiba and T.~Suzuki,
Phys.\ Lett.\ B {\bf 333} (1994) 461;\,
J.~D.~Stack, S.~D.~Neiman and R.~J.~Wensley,
Phys.\ Rev.\ D {\bf 50} (1994) 3399.
\bibitem{Suzuki:1989gp}
T.~Suzuki and I.~Yotsuyanagi,
Phys.\ Rev.\ D {\bf 42} (1990) 4257.
\bibitem{Luscher:2002qv}
M.~L\"uscher, P.~Weisz,
JHEP {\bf 0207} (2002) 049.
\bibitem{Kaczmarek:2000mm}
O.~Kaczmarek, F.~Karsch, E.~Laermann and M.~Lutgemeier,
Phys.\ Rev.\ D {\bf 62} (2000) 034021.
\end{thebibliography}
\end{document}